\documentclass[
 aps,
 prx,nofootinbib,
 amsmath,amssymb,
 reprint,%
]{revtex4-2}

\usepackage{graphicx}% Include figure files
\usepackage{dcolumn}% Align table columns on decimal point
\usepackage{xcolor}
\usepackage{bm}% bold math
\usepackage{upgreek}
\usepackage[utf8]{inputenc}
\usepackage[T1]{fontenc}
\usepackage{mathptmx}
\usepackage{etoolbox}
\usepackage{hyperref}
\usepackage{cleveref}
\usepackage{physics}
\usepackage{lipsum}

\usepackage[normalem]{ulem}
\usepackage{xcolor}  % allows text coloring

\newcommand\marklessfootnote[1]{
    \addtocounter{footnote}{-1} \renewcommand\thefootnote{}
    \footnotetext{#1}
}

\begin{document}

\title{Basic cell for a quantum microwave router}
% Force line breaks with \\
    \author{E.~Mutsenik\textsuperscript{\ddag}}
    \email{evgeniya.mutsenik@leibniz-ipht.de}

    \affiliation{Leibniz Institute of Photonic Technology, D-07745 Jena, Germany}
    % \altaffiliation{\textdagger\ These authors contributed equally to this work.}
    	
    \author{A.~Sultanov\textsuperscript{\ddag}}
\email{aidar.sultanov@leibniz-ipht.de}
	\affiliation{Leibniz Institute of Photonic Technology, D-07745 Jena, Germany}
    % \altaffiliation{\textsuperscript{\dag} }

    \author{L.~Kaczmarek}
	\affiliation{Leibniz Institute of Photonic Technology, D-07745 Jena, Germany}

    \author{M.~Schmelz}
	\affiliation{Leibniz Institute of Photonic Technology, D-07745 Jena, Germany}

    		\author{G.~Oelsner}
	\affiliation{Leibniz Institute of Photonic Technology, D-07745 Jena, Germany}
 
	\author{R.~Stolz}
	\affiliation{Leibniz Institute of Photonic Technology, D-07745 Jena, Germany}

        \author{E.~Il'ichev}
	\affiliation{Leibniz Institute of Photonic Technology, D-07745 Jena, Germany}

\begin{abstract}
We report the first experimental realization of a scalable basic cell for quantum routing, enabling coherent control and exchange of microwave photons between two spatially separated superconducting waveguides coupled via a single transmon qubit. The cell was characterized at 10~mK with an average input signal of $\approx$1~photon at $\approx$~6~GHz, and with the qubit biased to its optimal point to minimize sensitivity to external magnetic fluctuations. By combining steady-state and time-domain measurements, we reconstructed the key parameters of the system, including qubit relaxation and dephasing, waveguide–qubit couplings, and cross-waveguide photon transfer efficiency. The observed performance is consistent with a non-Hermitian Hamiltonian formalism and demonstrates clear limits set by flux bias, temperature, and photon number, in agreement with flux- and temperature-induced dephasing models. Crucially, the cell operates reliably at the single-photon level, and in the high-photon regime we directly observe photon dressing induced by the qubit. These results establish a versatile platform for studying open quantum system phenomena and pave the way for scalable implementations of quantum routing and network nodes.
\end{abstract}

\maketitle

\section{Introduction}

\marklessfootnote{\textsuperscript{\ddag} These authors contributed equally to this work.}
The implementation of quantum information processing devices requires scalable networks that connect  quantum processors via noiseless communication channels and provide controlled routing for information transfer \cite{Kimble2008}.  Here, photons are considered among the best candidates for carrying quantum information  as they can propagate with low loss and at high speed over long distances. Recent developments have led to a relatively new experimental platform for studying light-matter interaction, in which quantum emitters are coupled to propagating photons in a waveguide \cite{Sheremet2023,Hamann2018,Zanner2022,Brehm2021,Zanner2025,Kannan2023}. These systems, depending on their frequency range, exploit different realizations of microwave and optical waveguides coupled to artificial or natural atoms. Since here the photon dynamics are limited by the waveguide geometry, the obvious similarity with cavity QED \cite{Walther2006} or circuit QED \cite{Blais2021} has led to a similar terminology - this field is called waveguide quantum electrodynamics (WQED) \cite{Sheremet2023}.

The basic idea of the WQED device is to control the interaction of propagating photons with quantum emitters. Natural atoms are usually  microscopic in size, in this case, an ensemble of $N$-atoms is used to implement the optimal emitter-waveguide coupling  \cite{Sheremet2023}.  In contrast,   a "giant" atom - an artificial atom with a sufficiently large dipole moment or  spatially separated coupling ports- can provide an equivalent or even enhanced effective coupling \cite{Blais2021,Kockum2021}. 

Different implementations use different frequency ranges for quantum communication and information processing devices. In particular, the superconducting platform utilizes the microwave frequency domain. Consequently, controlled routing of microwave signals at cryogenic temperatures is required. More than a decade ago simple single-photon routing was experimentally demonstrated \cite{Delsing2011}: an input signal was transmitted to one of two output ports by controlling the electromagnetically induced transparency of the transmon. 

Recently, this direction has attracted a renewed attention \cite{Yan2014, Zhu2019, Gong2024} due to the potential implementation of hybrid optical-microwave quantum networks \cite{Bartholomew2020, Wang2018, Wang2022}, as well as the improvement of superconducting scalable quantum processors architecture \cite{Gong2024}. Developing in this direction requires the experimental realization of a simple, controlled quantum node that can be easily integrated into networks or processors.

Building on our previous theoretical proposals \cite{Sultanov2020}, herein we report on the first experimental realization of a minimal four-port single-photon router, implemented with a superconducting transmon coherently coupled to two independent waveguides. The conceptual sketch of this device is presented in Fig. \ref{fig:sketch}. 
This device serves as a fundamental building block that opens the way toward scalable implementations of coherent single-photon routing and enables the study of the qubit backaction on the propagating field, confirming key aspects of light–matter interaction in this platform.
The realization of such a controllable quantum node establishes a versatile building block for scalable quantum networks and modular quantum processors, providing an experimentally accessible system for both, practical applications and fundamental studies of open quantum systems, where the bandwidth is not limited by cavities or resonators.
The paper is organized as follows: In Sec. II, we discuss the theoretical approach to describe the basic cell's responses.  Section III is devoted to extraction of parameters of the basic cell, including calibration procedure of the steady-state responses and time-domain characterization.  We investigate the basic cell's performance under different experimental conditions as flux-bias dependence, photon number dependence and temperature dependence  in Section IV . Section V concludes the paper. 

\section{Theory}

\begin{figure}
    \centering
    \includegraphics[width=\linewidth]{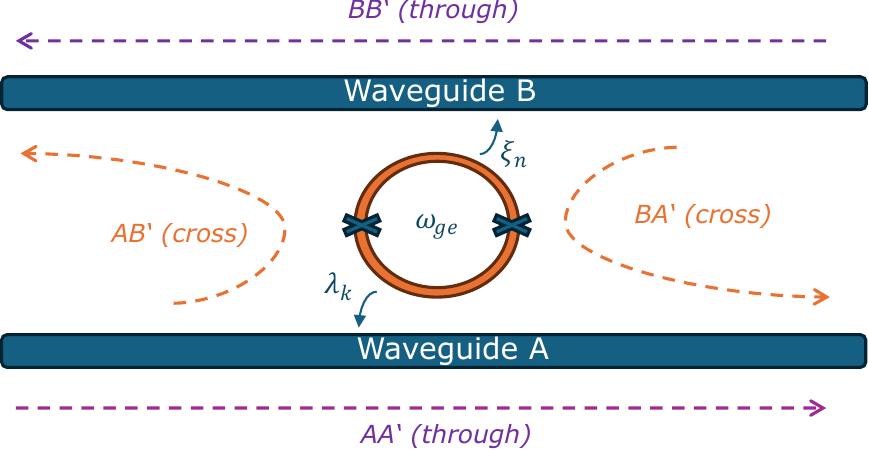}
    \caption{Conceptual sketch of the router basic cell.}
    \label{fig:sketch}
\end{figure}

In the basic cell for the router shown in Fig. \ref{fig:sketch}, the qubit is positioned between two open waveguides, and is coupled to both. Below, we consider the following possible photon transmission paths: 
\begin{enumerate}
    \item \textit{$\textbf{AA}^{\prime}$} \textbf{\textit{(through):}} input and output ports are in the waveguide~\textit{A};
    \item \textit{$\textbf{BB}^{\prime}$} \textbf{\textit{(through):}} input and output ports are in the waveguide~\textit{B};
    \item \textit{$\textbf{AB}^{\prime}$} \textbf{\textit{(cross):}} input port in the waveguide~A and output port in the waveguide~\textit{B};
    \item \textit{$\textbf{BA}^{\prime}$} \textbf{\textit{(cross):}} input port in the waveguide~B output port in the waveguide~\textit{A}.
\end{enumerate}

All the paths are schematically illustrated in Fig.~\ref{fig:sketch}.

Using a non-Hermitian Hamiltonian approach \cite{Greenberg2015}, we calculated the transmission coefficients in the same manner as it was done in \cite{Sultanov2020}. Below, we outline the main steps of the calculations, omitting details.

The Hamiltonian of the system under study reads:

\begin{widetext}
\begin{equation}
% \centering
    \hat{H} = \frac{\hbar}{2}\omega_{ge} \sigma_z + \hbar\sum\limits_{k} a^\dagger_{k} a_{k} \omega_{k} + \hbar \sum\limits_{n}b^\dagger_{n} b_{n} \omega_{n} + \hbar \sum\limits_{k}\lambda_{k}(a^\dagger_{k}\sigma_- + a_{k}\sigma_+) + \hbar \sum\limits_{n}\xi_{n}(b^\dagger_{n} \sigma_- + b_{n}\sigma_+),
\label{eq:Hamiltonian}
\end{equation}
\end{widetext}
where $\hbar = \frac{h}{2\pi}$ is the reduced Planck's constant, $\omega_{ge}$ is the qubit transition frequency between the ground and first excited states, $\sigma_z$ is the Pauli spin operator, $a^\dagger_{k} a_{k} (b^\dagger_{n} b_{n})$ are the creation/annihilation operators in the waveguide~\textit{A}(\textit{B}), ${\omega_{k(n)}}$ is the frequency of the \textit{k}-th (\textit{n}-th) mode in the corresponding waveguide, $\lambda_{k} (\xi_{n})$ is the coupling strength between \textit{k}-th (\textit{n}-th) mode in the corresponding waveguide~\textit{A}(\textit{B}) and the qubit, $\sigma_- (\sigma_+)$ is the lowering (rising) operator.

The Hamiltonian~(\ref{eq:Hamiltonian}) acts in a single-excitation Hilbert space, where the basis states can be expressed as $|\text{state}\rangle = |x,y,z\rangle$. Here $x(z)$ denotes the presence of the photon of the \textit{x}-th (\textit{z}-th) mode in the waveguide~\textit{A}(\textit{B}) respectively and \textit{y} represents the qubit state (excited $|e\rangle$ or ground $|g\rangle$).

Consequently, the basis for this case reads:

\begin{itemize}
    \item $|A_k\rangle = |1_k, g, 0_n\rangle$ - the photon is in the \textit{k}-th mode of the waveguide~\textit{A}, and the qubit is in the ground state. The total energy of this state is: $E_k= -\frac{\hbar}{2}\omega_{ge}+\hbar \omega_k$;
    
 \item $| 1\rangle = |0_k,e,0_n \rangle$ - the qubit is in the excited state, no photon in both waveguides;
 \item $|B_n \rangle = |0_k, g, 1_n\rangle$ - the photon is in the \textit{n}-th mode of the waveguide~\textit{B}, and the qubit is in the ground state. The total energy for this state is $E_n=-\frac{\hbar}{2}\omega_{ge}+\hbar\omega_n$; 

 \item $|C \rangle=|0_k,g,0_n\rangle$ -  no photon in both waveguides, the qubit is in the ground state.
\end{itemize}

To describe the photon dynamics defined by Hamiltonian~(\ref{eq:Hamiltonian}) we used the following procedure. In the first step, we partition the Hilbert space of (\ref{eq:Hamiltonian}) into two orthogonal subspaces, and define the corresponding projection operators:
\begin{equation*}
     P=\sum \limits_k  |A_k\rangle \langle A_k| +\sum \limits_n |B_n\rangle \langle B_n|,
\end{equation*}
which projects onto the continuum of states, and
\begin{equation*}
    Q=|1\rangle\langle1| + |C\rangle\langle C|
\end{equation*}
which projects onto the discrete states. We then project the full Hamiltonian~(\ref{eq:Hamiltonian}) onto these subspaces and eliminate the \textit{P}-subspace dynamics. This allows for effectively describing the photon scattering processes in the discrete subspace $Q$, while taking into account the influence of the eliminated subspace $P$. As a result, we obtain an effective Hamiltonian  $\hat{H}_{\text{eff}}$, which accounts for the influence of the dynamics of the entire system on the corresponding subspace. For more details see \cite{Greenberg2015}. In matrix form, $\hat{H}_{\text{eff}}$ it the diagonal matrix with the following elements:  

\begin{align}
   \langle C| \hat{H}_{\text{eff}}|C\rangle = -\frac{\hbar}{2}\omega_{ge}\\
   \langle1| \hat{H}_{\text{eff}}|1\rangle = \frac{\hbar}{2}\omega_{ge} - i\hbar\Gamma_A -i\hbar\Gamma_B,
\end{align}
where \textit{i} is the imaginary unit, $\Gamma_{A}=\frac{\lambda^2L}{ v}$ and $\Gamma_B=\frac{\xi^2L}{v}$ represent the coupling strengths of the qubit to the respective waveguides, which are associated with the half width at half maximum. These couplings are obtained under the assumption that the coupling strengths  are independent of the mode i.e.   $\lambda_k=\lambda, \xi_n=\xi$. Here $v$ is the group velocity in the waveguide, and $L$ is the length of the corresponding waveguide which is the same for both.      

Let's assume that the initial state is $|A_{0} \rangle = |1_{k_{0}}, g,0_n \rangle$, which means that at the beginning the incident photon is in the $k_0$ mode of the waveguide~A. %on the left side of the qubit.
The general representation of the wavefunction is:
\begin{equation}
    |\Psi_A \rangle = |A_{0} \rangle + \hbar^2\sum \limits_{k}\frac{  \lambda_{k}\lambda_{k_0} |A_k\rangle }{E_0-E_{k}+i\epsilon}R_{11} + \hbar^2\sum \limits_{n}\frac{\xi_{n}\lambda_{k_0} |B_n\rangle }{E_0-E_{n}+i\epsilon}R_{11},
\end{equation}
where $E_0$ is the energy of the initial state, $i\epsilon$ is a small imaginary addition preventing singularities, and $R_{11} = \langle1|\frac{1}{E_0-\hat{H}_{eff}}|1\rangle$ denotes the matrix element of the inverse of ${E_0-\hat{H}_{eff}}$. Now, consider the wave function in coordinate representation. For the photon at position $x_A$ in the waveguide~A the wavefunction is $\langle x_A|A_k\rangle = e^{ikx_A}$. Similarly to the photon in the waveguide~B at the position $x_B$: $\langle x_B|B_n\rangle=e^{inx_B}$.
The probability of detecting the photon in the waveguide~\textit{A} could be found as
\begin{equation}\label{eq:WF wvgA}
    \langle x_A|\Psi_A\rangle = e^{ik_0x_A} + \hbar^2\sum \limits_{k} \frac{e^{ikx_A}\lambda_k \lambda_{k_0}}{E_0 - E_{k}+i\epsilon}R_{11}.
\end{equation}
In the same way, the probability of detecting the photon in the waveguide~B is
\begin{equation}\label{eq:WF wvgB}
    \langle x_B|\Psi_A \rangle = \hbar^2 \sum \limits_{n}\frac{e^{inx_B}\xi_n \lambda_{k_0}}{E_0 - E_n + i\epsilon}R_{11}.
\end{equation}

Considering that the qubit placement position is $x=0$ , we can define the following transmission coefficients: $AA^\prime$ and $AB^\prime$, which indicate the side of the qubit where the incident photon is detected: left ($AB^\prime$) or right ($AA^\prime$), see Fig.\ref{fig:sketch}. Following the original method \cite{Greenberg2015} we replace the sum with the integration over all $k$ and $n$, taken by use of residue calculus with the poles located near $E_0$, i.e. $\hbar\omega_k\approx\hbar\omega_0$

\begin{equation}
    \langle x_A|\Psi_A \rangle = e^{ik_0 x_A}(1-i\hbar\Gamma_AR_{11}), 
\end{equation}
taking into account the explicit expression for $R_{11}$ we obtain
\begin{equation}
\label{eq:tAA'}
\langle x_A \mid \Psi_A \rangle 
= 
e^{i k_0 x_A} 
\underbrace{
\left(1 - \frac{i \hbar\Gamma_A }{\hbar \omega_{k} - \hbar \omega_{ge} + i \hbar \Gamma_A + i \hbar \Gamma_B} \right)}_{t_{AA^\prime}},
\end{equation}
where $t_{AA^\prime}$ is the transmission coefficient $AA^\prime$ \textit{through}. For the $AB^\prime$, the equation is as follows:
\begin{equation}
\label{eq:tAB'}
    \langle x_B|\Psi_A\rangle=-e^{ik_0x_B} \underbrace{\left(\frac{i\hbar{\sqrt{\Gamma_A \Gamma_B}}}{\hbar\omega_{k}-\hbar\omega_{ge}+i\hbar\Gamma_A+i\hbar\Gamma_B}\right)}_{t_{AB^\prime}}.
\end{equation}

On the other hand, for the incident photon in the waveguide~B with the initial state $|B_0\rangle=|0_k,g,1_{n_0}\rangle$, we can find transmission coefficients in the same way as was done for the state $|A_0\rangle$. For this case, the wavefunction is:

\begin{equation}
    |\Psi_B\rangle=|B_0\rangle+ \hbar^2\sum \limits_k\frac{\lambda_k\xi_0|A_k\rangle }{E_0-E_k+i\epsilon}R_{11}+ \hbar^2\sum \limits_n\frac{\xi_n \xi_0 |B_n\rangle }{E_0-E_n+i\epsilon}R_{11}.
\end{equation}
The probability that the photon is detected in the waveguide~A is
\begin{equation}
    \langle x_A|\Psi_B\rangle = \hbar^2\sum \limits_k\frac{e^{ikx_A}\lambda_k \xi_0}{E_0-E_k+i\epsilon}R_{11}
\end{equation}
and for the waveguide~B
\begin{equation}
    \langle x_B|\Psi_B\rangle = e^{in_0x_B} + \hbar^2\sum \limits_n \frac{e^{inx_B}\xi_n \xi_0}{E_0-E_n+i\epsilon}R_{11}.
\end{equation}
The transmission coefficients are thereafter:
\begin{equation}
\label{eq:tBB'}
    \langle x_B|\Psi_B\rangle = e^{in_0x_B}\underbrace{\left(1-\frac{i\hbar\Gamma_B}{\hbar \omega_n - \hbar \omega_{ge} +i\hbar \Gamma_A +i\hbar\Gamma_B}\right)}_{t_{BB^\prime}}
\end{equation}
and
\begin{equation}
\label{eq:tBA'}
    \langle x_A|\Psi_B\rangle =-e^{in_0 x_B}\underbrace{\left(\frac{i\hbar\sqrt{\Gamma_A \Gamma_B}}{\hbar \omega_n - \hbar\omega_{ge} +i\hbar\Gamma_A +i\hbar\Gamma_B} \right)}_{t_{BA^\prime}}.
\end{equation}

Accordingly, the steady-state transmission coefficients follow directly from the non-Hermitian Hamiltonian formalism used. If desired, this approach can be extended to obtain all possible sets of coefficients. Here, we consider only those coefficients that are relevant for comparison with the experimental data, as presented in the following chapters.

\section{Basic cell parameters}

In this section, we extract the main parameters that characterized our device, namely:
the coupling constants between qubit and the waveguides, the qubit lifetimes (both from steady state and time domain) and the residual coupling between the waveguides (effective isolation).  

\subsection{\label{sec:Sample introduction}Experimental setup}

\begin{figure}
    \centering
    \includegraphics[width=\linewidth]{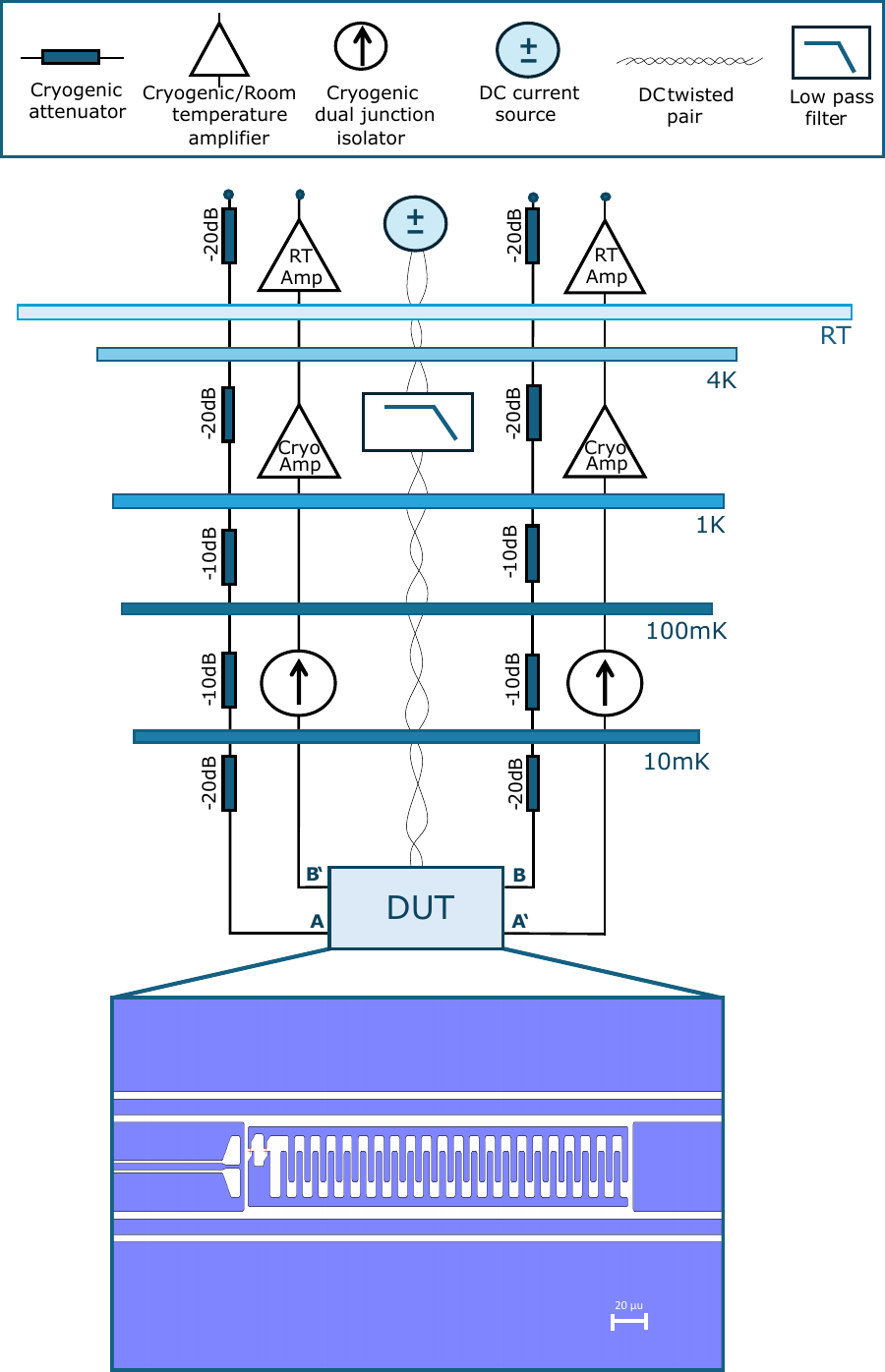}
    \caption{Sketch of the measurement setup.}
    \label{fig:setup}
\end{figure}

A sketch image of the central part of the sample is shown in Fig.\ref{fig:setup}.  
The qubit and coplanar open waveguides are made of aluminum on silicon substrate \cite{Schmelz2024}. The waveguides are designed identically with the following parameters: the central line width is 10~$\upmu$m and the gaps to ground planes are 5~$\upmu$m to ensure 50~$\upOmega$ impedance matching. A tunable transmon-type qubit \cite{Koch2007} fabricated with aluminum manhattan technology and with an integrated DC-bias control line was used. To suppress unwanted parasitic modes, aluminum air bridges were implemented
along the entire length of the waveguides \cite{Kaczmarek2025}. 

The sample was placed on the 10~mK temperature stage of a dilution refrigerator. A simplified sketch of the measurement setup is presented in Fig.\ref{fig:setup}. The input radio frequency  (RF) signal is attenuated by -20~dB at room temperature and by -70~dB in the dilution refrigerator. After passing through the sample, the signal is amplified on the 4~K temperature stage using a HEMT amplifier with gain of +40 dB and an additional amplifier with +20~dB gain at room temperature. All frequency domain transmission measurements were performed using a vector network analyzer (VNA) unless otherwise noted. All experiments were performed with all four ports connected to 50~$\upOmega$ sources or amplifiers to eliminate the appearance of additional reflections. 

\subsection{Transmission and isolation}
\begin{figure*}[htbp]
    \centering
    \includegraphics{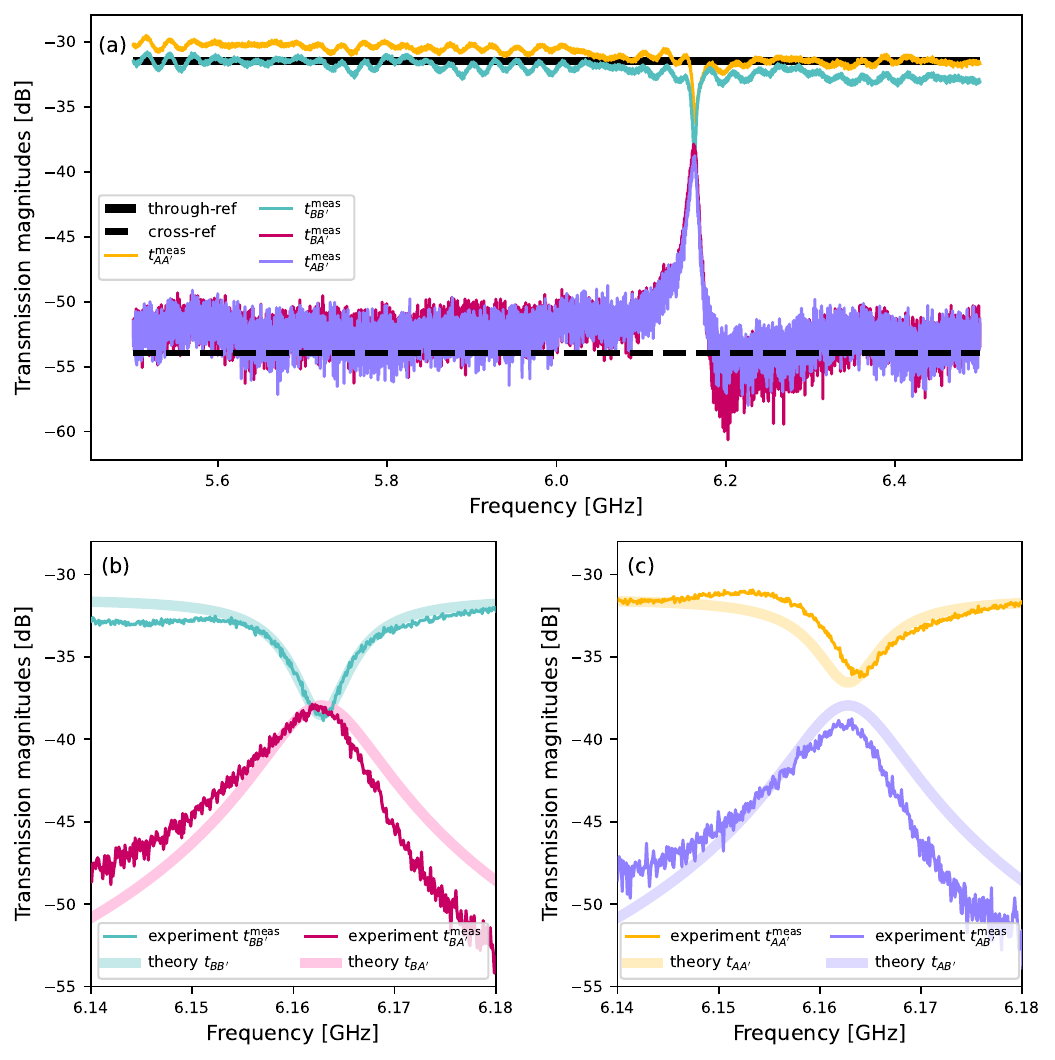}
    \caption{Measured transmission coefficients of the basic cell in comparison with the theory: a) measured transmission coefficients in a wide frequency bandwidth demonstrate the qubit response at the sweet spot. Black solid and dashed lines correspond to references curves (see details in the main text); b) transmission magnitude of the qubit response in a narrow frequency bandwidth ($t_{B B^{\prime}}^{\text{meas}}$, $t_{B A^{\prime}}^{\text{meas}}$) and comparison with the theoretical coefficients (Eq.~\ref{eq:tBB'}, Eq.~\ref{eq:tBA'}); c) transmission magnitude of the qubit response in a narrow frequency bandwidth ($t_{A A^{\prime}}^{\text{meas}}$, $t_{A B^{\prime}}^{\text{meas}}$) and comparison with the theoretical coefficients (Eq.~\ref{eq:tAA'}, Eq.~\ref{eq:tAB'}).  }
    \label{fig:transmission_coeffcients}
\end{figure*}

The raw transmission coefficients (output to input signals ratios) measured over a 1 GHz bandwidth are shown in Fig.~\ref{fig:transmission_coeffcients}~a, when the qubit is tuned to the sweet spot bias—i.e., the bias point at which the qubit is least sensitive to external magnetic field disturbances.
As expected and in line with \cite{Astafiev2010}, a dip is observed at the first qubit transition frequency $\omega_{ge}/(2 \pi) \approx 6.163$ GHz for the through transmissions ($AA'$ and $BB'$). The cross-transmissions ($A B^\prime$ and $B A^\prime$) have a peak at the same $\omega_{ge}$. This result is explained by photon re-emission into the waveguides after being excited. The input power is -140~dBm, which corresponds to an average of $\approx$~1 photon interacting with the qubit during its lifetime. Although the minimum distance between the waveguides is 68~$\upmu$m, the isolation between waveguide~A and waveguide~B is approximately $-20$~dB, estimated from the difference between corresponding reference curves ($\approx-30~\text{dB}$ for the through-transmission level  and $\approx-50~\text{dB}$ for the cross-transmission), as shown by black solid and dashed curves in Fig.~\ref{fig:transmission_coeffcients}~a. The reference curves represent the mean values of the respective magnitudes.  These results confirm that cross-transmission occurs \textit{mainly} via the qubit. 

The measured coefficients, however, do not directly reflect the intrinsic transmission properties of the basic cell. Instead, they represent effective responses of the entire measurement setup, which include the basic cell and the surrounding circuitry (such as connectors, attenuators, and impedance mismatches). As a result, the cell's behavior with various external contributions is measured. For example, the difference in background level between coefficients $A A^\prime$ and $B B^\prime$ is $\approx$~1~dB. We attribute this to slight differences between the coaxial lines in the dilution refrigerator, which may not be perfectly identical. To clearly distinguish between the coefficients of the basic cell and the measured ones, we denote the measured transmission coefficients with a superscript "meas", e.g.
\begin{equation*}
     t_{AA^\prime}^{\text{meas}},t_{BB^\prime}^{\text{meas}}, t_{AB^\prime}^{\text{meas}},t_{BA^\prime}^{\text{meas}}.
\end{equation*}

A comparison of the measured data with the theoretical predictions is shown in Figs.~\ref{fig:transmission_coeffcients}~b,c. For the transmission coefficients with the input in waveguide~B, the theoretical curves are plotted using Eqs.~\ref{eq:tBB'} and \ref{eq:tBA'}. Accordingly, for the waveguide~A as the input, the theoretical curves are given by Eqs.~\ref{eq:tAA'} and \ref{eq:tAB'}. In Fig.~\ref{fig:transmission_coeffcients}, the theoretical curves are simply rescaled to match the experimental signal level as a first order approximation to account for the aforementioned external contributions.

Overall, the theoretical curves are in reasonable agreement with the measured values. However, some discrepancy between the theoretical and experimental data is observed, see Figs.~\ref{fig:transmission_coeffcients}~b,c. Moreover, the obtained $t_{AA^\prime}^{\text{meas}}$ and $t_{BB^\prime}^{\text{meas}}$ exhibit slightly different behavior. 

Below we present a procedure that allows for self-consistent calibration of data and discuss the limitations of this method. 

\subsection{Steady-state response and data normalization}\label{sec:Calibration}

\begin{figure}[!h]
    \centering
    \includegraphics[width=\linewidth]{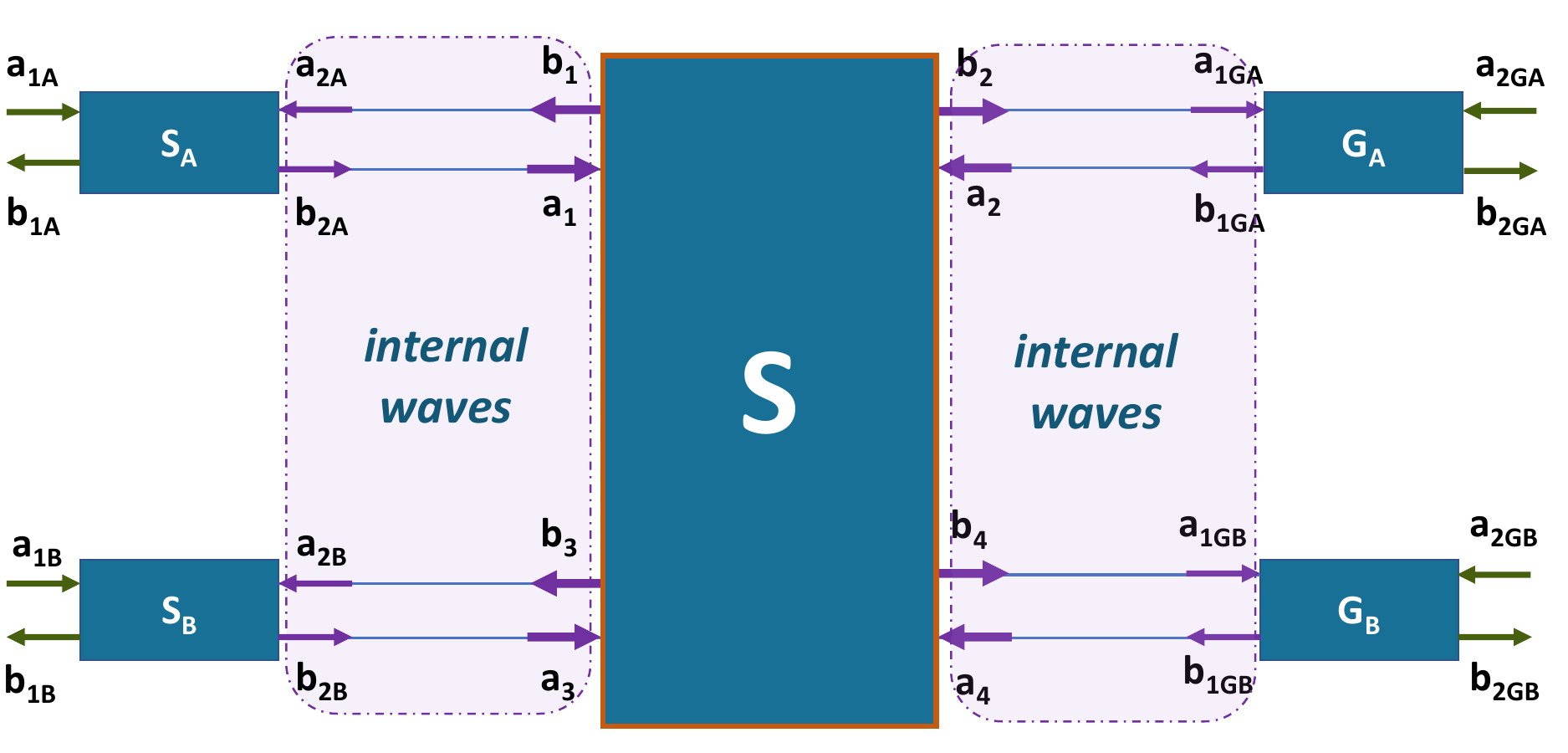}
    \caption{Equivalent multiport network circuit for the basic cell with input and output lines. Here, $a_i$ are the incoming waves to the $i$-th network, and $b_i$ are the outgoing waves. $S_{A/B}$ denotes the S-matrix of the input lines, $G_{A/B}$ that of the output lines, and $S$ is the S-matrix of the basic cell.}
    \label{fig:matrix_scheme}
\end{figure}

We aimed to reconstruct the internal transmission properties of the basic cell, isolating its intrinsic behavior from that of the measurement infrastructure. 
To achieve this reconstruction of the internal transmission properties and to compare them with our model, the following procedures must be performed:
\begin{enumerate}
    \item Calibration the raw data to quantify effects of input/output lines.
    \item Introducing small complex phases into the coupling coefficients to phenomenologically account for Fano-type interference and  impedance mismatches.
    \item Separation of the qubit intrinsic losses and losses defined by coupling to waveguides.
\end{enumerate}
%Below we demonstrate that this approach allows for a systematic interpretation of experimental data.
To eliminate an influence of the input and output lines, we model the setup using a combination of two-port scattering $S$-matrices for the input lines labeled as $S_{A/B}$ and output matrices labeled as $G_{A/B}$ and a four-port $S$-matrix for the basic cell (Fig.~\ref{fig:matrix_scheme}). 

%This allows us to isolate the response of the cell from line imperfections.
In Appendix \ref{app:Seffective} we derived an effective $S_{\text{eff}}$-matrix, which takes into account all elements. For simplicity, we assume that reflections from the measurement lines are negligibly small, and the isolation can be represented as a completely  symmetric matrix. In this case, the measured  responses can be approximated as follows: 
\begin{equation}\label{eq:simplified_transport}
\begin{aligned}
    &t_{AA^{\prime}}^{\text{meas}}=S_A^{21}\cdot t_{AA^{\prime}} \cdot G_A^{21},\\
    &t_{BB^{\prime}}^{\text{meas}}=S_B^{21}\cdot t_{BB^{\prime}} \cdot G_B^{21},\\
    &t_{AB^{\prime}}^{\text{meas}}=S_A^{21}\cdot \left(t_{AB^{\prime}} + I\right) \cdot G_B^{21},\\
    &t_{BA^{\prime}}^{\text{meas}}=S_B^{21}\cdot \left(t_{BA^{\prime}} + I\right) \cdot G_A^{21},\\
\end{aligned}
\end{equation}
where $S_{A/B}^{21}$ are the off-diagonal elements of the input lines matrices, $G_{A/B}^{21}$ are the off-diagonal elements of the output lines, and $I$ is the isolation between the transmission lines. The equations (\ref{eq:simplified_transport}) without a qubit are:
\begin{equation}\label{eq:simplified_no_qb_transport}
\begin{aligned}
    &t_{AA^{\prime}}^{HD}=S_A^{21}\cdot  G_A^{21},\\
    &t_{BB^{\prime}}^{HD}=S_B^{21}\cdot G_B^{21},\\
    &t_{AB^{\prime}}^{HD}=S_A^{21}  \cdot I \cdot G_B^{21},\\
    &t_{BA^{\prime}}^{HD}=S_B^{21} \cdot I \cdot G_A^{21},\\
\end{aligned}
\end{equation}
To realize the applicability of Eqs. (\ref{eq:simplified_no_qb_transport}), a high input power was used to effectively decouple the qubit from the transmission lines. The according responses here are denoted as ${HD}$ (\textit{high drive}).  Note, that from Eqs. (\ref{eq:simplified_no_qb_transport}) the isolation is expressed as
\begin{equation}
I= \sqrt{\frac{t_{AB^{\prime}}^{HD} t_{BA^{\prime}}^{HD}}{t_{AA^{\prime}}^{HD} t_{BB^{\prime}}^{HD}}},
\label{eq:isolation}
\end{equation}
resulting in -20~dB, a value which is  close to the designed one and estimated from the raw data. 

Thus, Eqs. (\ref{eq:simplified_transport}-\ref{eq:simplified_no_qb_transport}) can be used to characterize the measurement lines and reconstruct the pure response of the basic cell.
Combining these equations yields the basic cell's responses: 
\begin{equation}\label{eq:calibrated_responses}
\begin{aligned}
t_{AA^{\prime}}=\frac{t_{AA^{\prime}}^{\text{meas}}}{t_{AA^{\prime}}^{HD}},\\
t_{BB^{\prime}}=\frac{t_{BB^{\prime}}^{\text{meas}}}{t_{BB^{\prime}}^{HD}},\\
t_{AB^{\prime}} =\left( t_{AB^{\prime}}^{\text{meas}}-t_{AB^{\prime}}^{HD} \right)\cdot \sqrt{\frac{t_{BA^{\prime}}^{HD}}{t_{AA^{\prime}}^{HD} \cdot t_{BB^{\prime}}^{HD}\cdot t_{AB^{\prime}}^{HD}}},\\
t_{BA^{\prime}} =\left( t_{BA^{\prime}}^{\text{meas}}-t_{BA^{\prime}}^{HD} \right)\cdot \sqrt{\frac{t_{AB^{\prime}}^{HD}}{t_{AA^{\prime}}^{HD} \cdot t_{BB^{\prime}}^{HD}\cdot t_{BA^{\prime}}^{HD}}},\\
\end{aligned}
\end{equation}
Prior to using Eq.~\eqref{eq:calibrated_responses} to reconstruct the intrinsic responses of the basic cell, the measured data are corrected for global phase offsets and phase discontinuities. Since taking the square root of complex signals halves their phase, discontinuities of $2\pi$ in the original data appear as $\pi$ jumps. To handle these correctly, we apply a modified unwrapping procedure: the phase angles are first doubled, unwrapped using a standard algorithm, and then halved to return to the original scale. A global phase shift is then removed such that the phase at resonance is zero, ensuring consistency with measurement conventions. These corrections do not alter the physical content of the data, providing a well-defined, continuous phase suitable for comparison with theoretical models.
%and the details of the phase adjustment procedure is given in the Supplementary$.  

\begin{figure*}
    \centering
    \includegraphics[width=\linewidth]{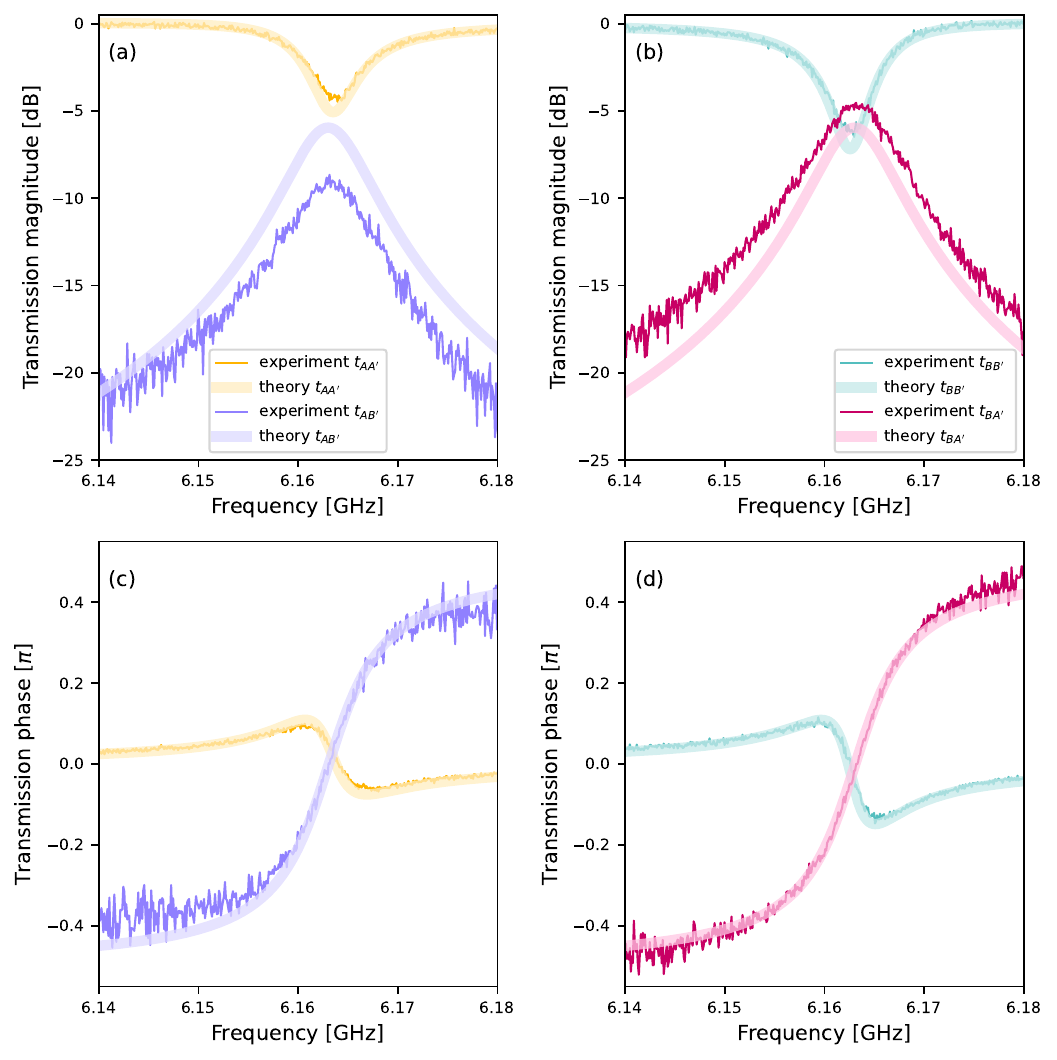}
    \caption{Calibrated transmission coefficients with fits. $\Gamma_A=2\pi \cdot 1.82~\text{MHz},\Gamma_B=2\pi \cdot 2.31~\text{MHz},\omega_{ge}=2\pi \cdot 6.163~\text{GHz},\varphi_A=-0.06\pi,\varphi_B=0.05\pi$.}
    \label{fig:calibrated_responses}
\end{figure*}

The experimental responses, calibrated in this way,  similar to the raw data shown in Fig. \ref{fig:transmission_coeffcients}, are presented in Fig. \ref{fig:calibrated_responses}.

In the next step, we fit all four calibrated transmission coefficients simultaneously, as they share the same set of parameters $\Gamma_A$, $\Gamma_B$, and $\omega_{ge}$. To account for remaining deviations from the ideal lineshape, including Fano-type interference and minor impedance mismatches in the waveguides, we allow the coupling parameters $\Gamma_A$ and $\Gamma_B$ to acquire a small complex phase: $\Gamma_A e^{i\varphi_A}$ and $\Gamma_B e^{i\varphi_B}$. This phenomenological modification captures the leading-order effects of non-ideal transmission lines on the measured complex transfer functions which are not catched by the previous procedure, while keeping the total number of fit parameters minimal. The fitted curves, obtained with theoretical equations (\ref{eq:tAA'},\ref{eq:tAB'},\ref{eq:tBB'},\ref{eq:tBA'}) are shown in Fig.~\ref{fig:calibrated_responses}. The fitted phases, $\varphi_A \approx -0.06\pi$ and $\varphi_B \approx 0.05\pi$, are small, justifying the interpretation of the couplings as effectively real for the main physics, with the complex part serving purely as a minor phenomenological correction. The extracted coupling rates are $\Gamma_A=2\pi\cdot 1.82$~MHz, $\Gamma_B=2\pi\cdot 2.31$~MHz and $\omega_{ge}=2\pi\cdot 6.163$~GHz.

Residual discrepancies arise from the simplified model and calibration assumptions. In particular, internal qubit losses are not included in the original model, which can lead to over/under-estimate $\Gamma_{A/B}$. To separate the qubit losses and coupling losses, we perform circle fits on $t_{AA'}$ and $t_{BB'}$ \cite{Probst2015}, extracting loaded ($\kappa_L$)  rates: 
\begin{align*} \kappa_L^{AA'},\kappa_L^{BB'} \end{align*}
that correspond to the full width at half maximum. 
Since both responses are obtained from scattering on the same object (the qubit), we can formulate balance equations for the corresponding loss rates:
\begin{align}
\kappa_L^{AA'}=\kappa_i+2\Gamma_A+2\Gamma_B\label{eq:loaded_AA};\\
\kappa_L^{BB'}=\kappa_i+2\Gamma_A+2\Gamma_B. \label{eq:loaded_BB}
\end{align}

\begin{itemize}

    \item Equation \eqref{eq:loaded_AA} indicates the full energy loss rate when observing through the waveguide \textit{A}. Partially it is lossed in qubit itself $\kappa_i$, the rest is emitted to two waveguides. Similarly, equation \eqref{eq:loaded_BB} shows the full loss rates when observed through the waveguide \textit{B}. We obtain $\kappa_L^{AA'}=2\pi\cdot9.33$~MHz and  $\kappa_L^{BB'}=2\pi\cdot8.48$~MHz, which coincide within 10 \%, attributed to precision of quality factor extraction by the circle-fit method. Thus, we take the mean loaded rate as $\kappa_L=2\pi \cdot 8.9\pm 0.89$~MHz.

    \item Then, we can directly estimate the loss rate of the qubit as $\kappa_i=2\pi\cdot 0.64$~MHz. We would like to stress, that this loss rate is not purely energy related, i.e. one can't directly say that it directly maps to the qubit relaxation rate or its dephasing. Nevertheless, this estimation shows that the observed steady-state responses are mainly defined by the interaction between the qubit and the waveguides. 

\end{itemize}
 As the estimation of the quality factors could be affected by many factors \cite{Lupascu2013,Rieger2023},  this intrinsic loss rate is only a rough estimation of the real intrinsic losses of the qubit. Especially, in our case, the ratio of extracted rates, indicates that the system is in almost critically coupled regime, where the separation of different loss channels is not very robust.

While not being a complete system analysis, our approach provides a reliable framework for interpreting the experimental data. The model accurately reproduces the lineshape and frequency dependence of four transmission paths. However, the model does not capture the observed discrepancy in transmission magnitudes between \( t_{AB'} \) and \( t_{BA'} \) at resonance, which are expected to be equal due to reciprocity in the model. This asymmetry in the measured data is likely caused by reflections, impedance mismatches, or unmodeled internal losses, which are not included in the present framework. A more comprehensive model including these effects is left for future work.

\subsection{Time domain measurement}
Here we present the time-domain characterization of the basic cell. Following the method described in \cite{Sultanov2025}, the qubit dynamics were measured using transitions of a transmon embedded in open waveguides. Namely, we probe the population of the first excited state$|e\rangle$ by sending readout pulse probing the transition to the second excited state $|ef\rangle$. In contrast to the steady-state setup, where responses were measured using VNA, here we use a heterodyne pulsed setup implemented with the OPX+Octave system from Quantum Machines. The details of the system incorporation in QED experiments could be found at  \cite{Liu2024,Hutin2024}. All measurements were performed at the optimal flux bias point (sweet spot), where the transition frequencies are $\omega_{ge}  = 2\pi \cdot 6.163$~GHz and $\omega_{ef} = 2\pi \cdot 6.015$~GHz.
Results of three types of experiments are presented to demonstrate the time-domain control and relaxation properties of the system:

(a) Rabi oscillations at the $\ket{g}\leftrightarrow\ket{e}$ transition as a function of drive amplitude, used to calibrate $\pi$-pulses;

(b) Rabi oscillations versus pulse duration, used to estimate coherence times;

(c) energy relaxation measurements ($T_1$).

\begin{figure}[!hb]
    \centering
    \includegraphics[width=1\linewidth]{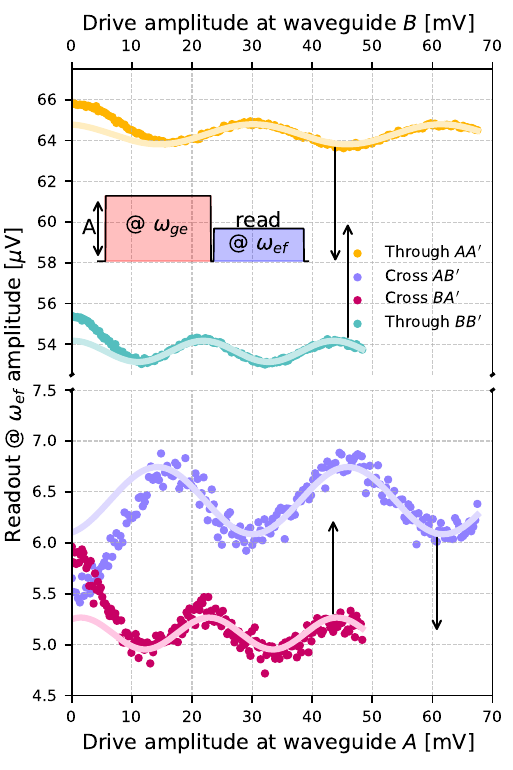}
    \caption{Rabi oscillation experiment with varying drive pulse amplitude A. The drive pulse is sent to one of the two waveguides at $|g\rangle \rightarrow|e\rangle$-transition frequency, the readout pulse is sent to the same waveguide as the drive pulse, but detected either there (through) either in opposite waveguide's output port (cross). Two X-axes are shown: the top axis corresponds to amplitudes applied to waveguide \textit{B}, and the bottom axis to waveguide \textit{A}. Arrows indicate which trace corresponds to which drive amplitude axis.  The data is shown in raw format, obtained with a standard heterodyne detection scheme. We note that the readout signal levels are different for cross and through measurements, resulting into the gap in the Y-axis.}
    \label{fig:raw_rabi_amplitudes}
\end{figure}

Firstly, $\pi$-pulse amplitudes are calibrated for both waveguides. For this purpose, 24~ns drive pulses with varying amplitude are applied to waveguides \textit{A} and \textit{B}, followed by a 60~ns readout pulse resonant with the $\omega_{ef}$ transition. The resulting signals are recorded either in the same waveguide as the drive one (“through” detection) or in the opposite waveguide (“cross” detection). The raw measurement data are presented in Fig.~\ref{fig:raw_rabi_amplitudes}, characterizing the important features of the basic cell operation:
\begin{itemize}
\item \textbf{DC offsets in raw data}: Raw data is obtained by down-converting the signal using a local oscillator (LO). If the readout pulse, formed by up-conversion using the same LO, contains a residual LO tone (due to incomplete suppression), this component appears as a DC offset in the measured signal.  Despite careful calibration of both the IQ mixers and the ADC offsets, a small leakage remains. Due to the finite isolation between waveguides \textit{A} and \textit{B}, this leakage is strongly suppressed in the cross-channel: the offset in the “through” signal is roughly an order of magnitude larger than in the “cross” signal, which is consistent with the independently measured isolation of –20 dB. In practice, such isolation acts as an effective filter suppressing technical noise, and this feature will be utilized in the following sections.
\item \textbf{Deviation from a simple Rabi fit under weak drive}: At low drive amplitudes, when the qubit remains predominantly in the $\ket{g}$ state, the measured response deviates from a perfect sinusoidal Rabi fit. This deviation arises because only a small fraction of the population is transferred to the $\ket{e}$ state, and thus the readout pulse interacts with the qubit only weakly. In this regime, the detected signal is determined primarily by the technical background rather than quantum scattering. Moreover, the duration of both excitation and readout pulses is comparable to the qubit lifetime, which enhances this effect.
\item \textbf{Drive amplitude and qubit rotations}: The final qubit state after a drive pulse depends on its amplitude, which creates the expected Rabi oscillations between $\ket{g}$ and $\ket{e}$, as shown in Fig.~\ref{fig:raw_rabi_amplitudes}. Due to the different coupling between the waveguides and the qubit, the required amplitude for $\pi$-rotation differs for the two waveguides: 
%driving through waveguide \textit{B} requires a lower amplitude than through waveguide \textit{A}. 
As a result of calibration, it was found that the ratio of the amplitudes of the $\pi$-pulses in the waveguides \textit{A} and \textit{B} is $\pi_A / \pi_B \approx 1.32$.  This result is in in good agreement with the independently obtained coupling ratio $\Gamma_B / \Gamma_A \approx 1.27$
from the steady-state analysis. This agreement confirms that the time-domain calibration is consistent with the frequency-domain characterization.
\end{itemize}

\begin{figure}[!ht]
    \centering
    \includegraphics[width=1\linewidth]{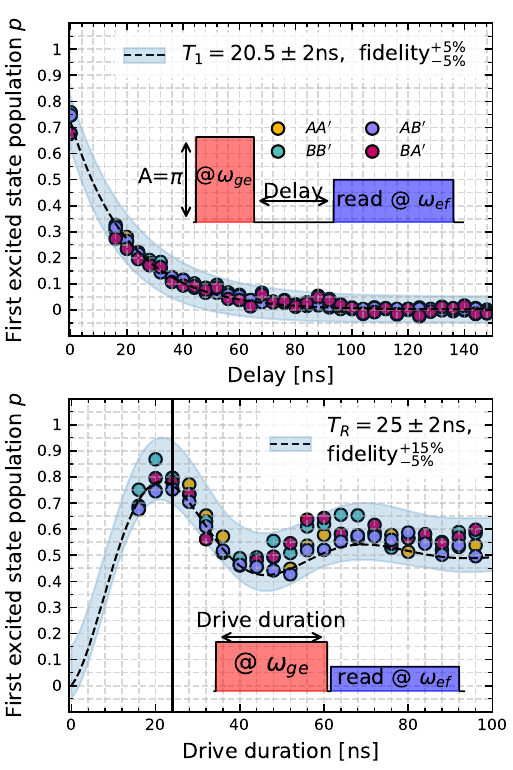}
    \caption{The energy relaxation  (top) and decay of Rabi-oscillation (bottom) with the reconstructed the first excited state population measured for four combinations of drive and readout waveguides.  The vertical line designates the drive pulse duration chosen for relaxation experiment and for $\pi$-pulse amplitude calibration.}
    \label{fig:decomposed_lifetimes}
\end{figure}

With the $\pi$-pulse amplitudes calibrated, 
the qubit’s temporal dynamics was investigated.  
To get optimal contrast of the output signal we reduce the duration of the readout pulse to 32~ns. Details of these measurements are presented in Appendix \ref{app:raw_time_domain}. 
The output signal reflects information about the qubit state populations. Essentially, each measurements result contains contributions from the two basic qubit states $\ket{g}$ and $\ket{e}$.
\begin{equation*}
    I+jQ=p(I_e+jQ_e)+(1-p)(I_g+jQ_g)
\end{equation*}

where $I(Q)_{g/e}$ are quadratures of the output signal for the qubit being in the ground or excited state and $p$ is the population of the excited state. Due to the finite qubit lifetimes the pure excited 
state signal $(I_e+jQ_e)$ is difficult to measure directly. To reconstruct it, similar to \cite{Biancetti2009,Biancetti2010}, we employ the following procedure:

\begin{itemize}
  \item  For each available drive and readout waveguides combination, we prepare qubit in different states using drive pulses of varying duration and amplitude. The results of subsequent measurements are displayed on the complex plane.
  \item  By making use of principal component analysis \cite{jolliffe2002pca}, we determine the axis with the highest variance, corresponding to changes in the qubit population. Thus, only one component of the output signal ($I_{g/e}$) can be used, after aligning the complex plane with the found axis.
  \item   The ground-state reference $I_g$ is fixed using the measurement with zero drive amplitude.
  \item  The excited-state reference $I_e$ is optimized such that all measured values can be decomposed linearly into populations $0\leq p \leq1$, allowing for a worst-case population error of 15\%.
  
 \item  This procedure is repeated for all drive readout waveguide combinations independently.
\end{itemize}
The applied procedure above results in the following: the population dynamics from both Rabi oscillation decay and energy relaxation experiments can be plotted on a unified scale for all drive–readout combinations, see Fig.\ref{fig:decomposed_lifetimes}.

The energy relaxation was fitted by an exponential decay:
\begin{equation*}
    p=p_0e^{-\frac{t}{T_1}}+p_{\infty}
\end{equation*}
where $p_0=0.72\pm 0.05$ is the population at the end of drive pulse,with uncertainty $\delta p_0=0.05$ accounting for state-preparation and readout infidelities. The offset  $p_{\infty}=1.6\cdot10^{-3}$ represents the residual excited state population at large delay times (see Fig.~\ref{fig:decomposed_lifetimes}), which arises due to finite measurement error and a low thermal population of the excited state. The relaxation time extracted from the fit is $T_1=20.5\pm 2$~ns, where the quoted uncertainty reflects the spread of $T_1$ values obtained from four independent measurements. In Fig.~\ref{fig:decomposed_lifetimes}, the average decay curve (using $T_1 = 20.5$~ns) is shown as black-dashed line together with two additional curves corresponding to $T_1 \pm 2$~~ns, forming a region between them (shown as a shaded band) encompasses all experimental data points.

The decay of driven Rabi oscillations was fitted using the expression:
\begin{equation*}
    p(t) = \left[p_{\max}\sin^2\!\left(\frac{\pi t}{2t_{\pi}}\right) - p_{\infty}\right]
    e^{-t/T_R} + p_{\infty},
\end{equation*}
where $p_{\max} = 1.2$ is the maximum oscillation amplitude, $t_{\pi} = 24~\text{ns}$ is the $\pi$-pulse duration, $p_{\infty} = 0.5$ is the residual excited-state population at long drive times, and $T_R = 25 \pm 2~\text{ns}$ characterizes the relaxation of the oscillation envelope. 

The average curve, corresponding to $T_R = 25~\text{ns}$, is shown by the black-dashed line in the bottom panel of Fig.~\ref{fig:decomposed_lifetimes}. The shaded error region represents the envelope of possible curves formed by varying both, the decay time within its uncertainty range ($T_R \pm 2$~ns) and the population offset with asymmetric error bounds $\delta p_\infty$  of $+0.15$ and $-0.05$, in contrast with the relaxation experiment. This combined envelope completely encompasses the experimental data points, capturing the total uncertainty in the measurement and fitting.

We aim to quantify the extent to which the observed qubit dynamics are determined by its designed coupling to the waveguides $\Gamma_A/\Gamma_B$, as opposed to other loss mechanisms such as intrinsic relaxation and dephasing. The extracted decay times are found to be consistent with the expected coupling strengths. In the energy relaxation experiment the qubit is sensitive to three different decay channels: waveguide~$A$ with a decay rate of $2\Gamma_A$, waveguide~$B$ with a decay rate of $2\Gamma_B$, and an "internal loss" channel into the thermal bath with a rate of $\Gamma_{bath}$. Accordingly, the energy relaxation rate is a linear combination of these  processes:
\begin{equation*}
    \Gamma_1=\frac{1}{T_1}=2\Gamma_A+2\Gamma_B+\Gamma_{bath}
\end{equation*}
An independent estimate of the relaxation time due to coupling to the waveguides gives
$\frac{1}{2\Gamma_A + 2\Gamma_B} \approx 19.3$~\text{ns}, which is in good agreement with the measured $T_1$, revealing that $\Gamma_{bath} \ll \Gamma_A,\Gamma_B$. 
Having established that energy relaxation is dominated by the engineered couplings, we now turn to the role of dephasing. For that,we decompose the decay rate of driven Rabi oscillations as combination of energy relaxation rate and pure dephasing rate $\Gamma_{\varphi}$, which works for the single qubit decay model \cite{Ithier2005}:
\begin{equation*}
    \Gamma_R=\frac{1}{T_R}=\frac{3\Gamma_1}{4}+\frac{\Gamma_{\varphi}}{2}
\end{equation*}
that allows us to estimate that the coupling to the thermal bath be is negligibly small and the pure dephasing rate ranges from $2\pi\cdot 0.94$~MHz to $2\pi\cdot3.2$~MHz, assuming the uncertainties in the rates estimations.This results in a pure dephasing rate that is 0.03-0.9 of the coupling rates. Although these ratios appear relatively large, they are derived under experimental constraints on state preparation and rely on simplifying assumptions from standard cQED models where the qubit is embedded in a resonator. However, the results demonstrate that the observed dynamics is predominantly governed by the engineered qubit–waveguide couplings. This consistency between steady-state and time-domain measurements confirms that the qubit and the two waveguides form a single coherently coupled quantum system. Such behavior justifies viewing this configuration as an elementary building block, or basic cell, for scalable WQED architectures.

Summarizing the steady-state measurements results and time domain characterization, we conclude that they are completely consistent. Moreover, energy relaxation and Rabi decay experiments  demonstrate that both waveguides behave predictably and yield stable results in all measurement configurations. These findings provide a solid experimental basis for further studies of the basic cell performance.

\section{Dependence of cell performance on external parameters.}
In this section, we analyze how the qubit-mediated interaction between waveguides depends on the flux-bias, the average number of input photons, and temperature.  It is also demonstrated that the proposed basic cell can serve as a platform for explorations of other fundamental effects.

\subsection{Flux-bias dependence or tunability}\label{sec:dephasing_due_to_flux}
Photon transfer between two waveguides is performed by a qubit. Therefore qubit dephasing is expected to reduce the transfer efficiency. The experiments described above were conducted in the sweet spot. In this regime, the pure dephasing rate is only weakly sensitive to flux fluctuations. Thus, the dephasing rate can be increased by moving away from the sweet-spot.

To test this expectation, we measured the steady-state transmissions as a function of the current applied to the integrated DC-bias coil coupled to the qubit. Leveraging the calibration protocol presented previously, we show below that the ratio of raw measured traces, following the approach in section \ref{sec:Calibration}, provides a convenient and experimentally accessible measure of photon transfer degradation due to qubit dephasing. From (\ref{eq:simplified_transport}) one obtains:
\begin{widetext}
\begin{equation} 
\label{eq:our_efficiency_def}
    \frac{        t_{AB^{\prime}}^{\text{meas}} t_{BA^{\prime}}^{\text{meas}}}{t_{AA^{\prime}}^{\text{meas}}    t_{BB^{\prime}}^{\text{meas}}}=
    \frac{        t_{AB^{\prime}} t_{BA^{\prime}}}{t_{AA^{\prime}}    t_{BB^{\prime}}}+I\left(\frac{        t_{AB^{\prime}} }{t_{AA^{\prime}}    t_{BB^{\prime}}}+\frac{       t_{BA^{\prime}}}{t_{AA^{\prime}}    t_{BB^{\prime}}} \right)+\frac{I^2}{t_{AA^{\prime}}    t_{BB^{\prime}}}\\ \approx  \frac{        t_{AB^{\prime}} t_{BA^{\prime}}}{t_{AA^{\prime}}    t_{BB^{\prime}}},
\end{equation}
\end{widetext}
where the last approximation is valid, because the isolation $I$ is low for our device (recall -20 dB from the section \ref{sec:Calibration}). 

For the single qubit mediator, dephasing can be incorporated phenomenologically by introducing an imaginary component to the qubit transition frequency:

\begin{equation*} 
\omega_{ge} \rightarrow \omega_{ge} - i \Gamma_\varphi,
\end{equation*}
In general, a common practice in cQED is to consider a damping via such imaginary terms \cite{Dalibard1992,Molmer2013,Azcona2025}. This substitution captures the leading-order effect of dephasing on steady state photon transfer, without requiring a full dynamical model of the qubit.
We assume that away from the sweet spot, the primary effect of flux biasing is to increase $\Gamma_\varphi$, with negligible influence from additional noise sources such as two-level fluctuators within the measured bandwidth. Substituting $\omega_{ge} \rightarrow \omega_{ge} - i \Gamma_\varphi$ into Eqs.~\ref{eq:tAA'}–\ref{eq:tBA'} yields:

\begin{widetext}
%\begin{gather}
\begin{equation}
  E =\frac{t_{AB^{\prime}}^{\text{meas}} \, t_{BA^{\prime}}^{\text{meas}}}
       {t_{AA^{\prime}}^{\text{meas}} \, t_{BB^{\prime}}^{\text{meas}}}=
  \\
    -\frac{\Gamma_A \Gamma_B}
    {\Delta^2 + i \Delta(\Gamma_A + \Gamma_B - 2\Gamma_\varphi)
    - \Gamma_\varphi (\Gamma_A + \Gamma_B + \Gamma_\varphi)
    - \Gamma_A \Gamma_B},
%\end{gather}
\end{equation}
\end{widetext}
where $\Delta=\omega - \omega_{ge}$ is the detuning from the qubit transition frequency, and $\omega\equiv\omega_k\equiv\omega_n$ is the frequency of the signal.
At resonance ($\Delta = 0$), this simplifies to:

\begin{equation}
\label{eq:transfer_efficiency_with_dephasing} 
E = \frac{t_{AB'}^{\rm meas} t_{BA'}^{\rm meas}}{t_{AA'}^{\rm meas} t_{BB'}^{\rm meas}} = \frac{1}{1 + \Gamma_\varphi \left(\frac{1}{\Gamma_A} + \frac{1}{\Gamma_B}\right) + \frac{\Gamma_\varphi^2}{\Gamma_A \Gamma_B}}.
\end{equation}

In the absence of dephasing ($\Gamma_\varphi = 0$), ideal coherent transfer is recovered ($E = 1$). As Eq.~\ref{eq:transfer_efficiency_with_dephasing} shows, $E$ decreases monotonically with increasing $\Gamma_\varphi$. When $\Gamma_\varphi \sim \Gamma_{A,B}$, the transfer efficiency is significantly reduced, with $E \approx 0.25$ in the strongly dephased regime.

\begin{figure} 
\centering \includegraphics[width=1\linewidth]{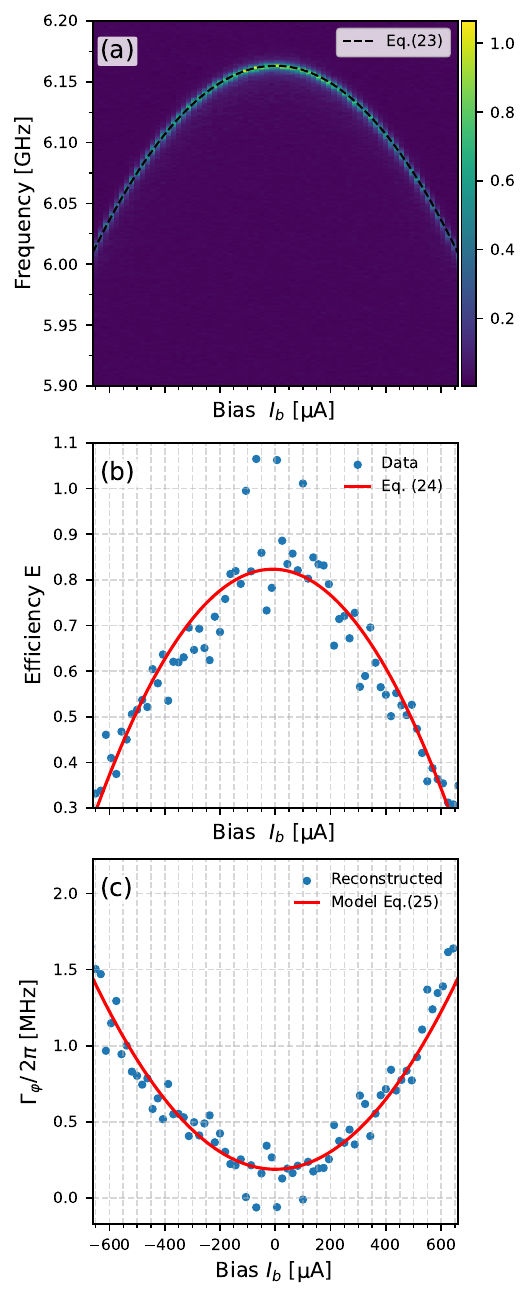} 
\caption{Influence of pure dephasing on photon transfer.
(a) Dependence of $E = \frac{t_{AB'}^{\rm eff} t_{BA'}^{\rm eff}}{t_{AA'}^{\rm eff} t_{BB'}^{\rm eff}}$ on flux bias current, where $I_b = 0$ corresponds to the transmon’s sweet spot. The black dashed line represents the polynomial fit for $\omega_{ge}$, see Eq.\ref{eq:wge_polinom}.
(b) Values of $E$ at the qubit frequency, showing a drop from the ideal value $E=1$ down to $E \approx 0.3$. The solid red line is the second-order polynomial fit used to reconstruct the pure dephasing rate.
(c) Reconstructed pure dephasing rate as a function of flux bias current. The solid red line represents the fit based on \cite{Ithier2005}.} 
\label{fig:iso_miso_bias} 
\end{figure}

Figure~\ref{fig:iso_miso_bias}~a shows the measured dependence of $E$ on flux bias current and signal frequency.  The maximum transfer occurs at the sweet spot, which we rescale to $I_b = 0$. As the bias is tuned away from this point, $E$ drops to approximately 0.5, confirming that the transfer process is mediated by qubit coherence. 
The peak position of the $E$ as a function of bias  dependence follows the same polynomial as the qubit frequency:
\begin{equation}\label{eq:wge_polinom}
\frac{\omega_{ge}(I_b)}{2\pi} = -352~\mathrm{MHz}/\mathrm{mA}^2 \cdot I_b^2 + 6.163~\mathrm{GHz},
\end{equation}
supporting the interpretation that dephasing is flux-noise-induced.

We extract the values of $E$ at $\Delta = 0$ and fit them with a second-order polynomial:
\begin{equation} 
\label{eq:Efficiency_polynomial}
E(I_b) = -1.29~{\rm mA}^{-2} \cdot I_b^2 - 0.025~{\rm mA}^{-1} \cdot I_b + 0.82.
\end{equation} 
This quantitative behavior confirms that the transfer is mediated by qubit coherence, as the quadratic suppression is consistent with Eq.~\ref{eq:transfer_efficiency_with_dephasing}.

Solving Eq.~\ref{eq:transfer_efficiency_with_dephasing} for $\Gamma_\varphi$, and taking only positive $\Gamma_\varphi>0$ solutions, we reconstruct the pure dephasing rate as a function of flux bias. The resulting values are shown as dots in Fig.~\ref{fig:iso_miso_bias}~c. We then fit the reconstructed dephasing using the standard expression for flux-noise-limited dephasing \cite{Ithier2005}:
\begin{equation}
\Gamma_\varphi = \pi \left( \frac{d\omega_{ge}(I_b)}{dI_b} \right)^2 S_I + \Gamma_\varphi^0, 
\end{equation} 
where $S_I$ is the spectral density of bias current noise and $\Gamma_\varphi^0$ accounts for dephasing from other sources. 

From the fit, we extract $S_I = 3 \times 10^{-19}~\mathrm{A^2/Hz}$ and $\Gamma_\varphi^0=2\pi\cdot 0.2$~MHz. This result provides a quantitative link between photon transfer efficiency and qubit coherence, offering a direct measure of flux-noise-induced decoherence in our system.

\subsection{Impact of thermal fluctuations on the basic cell perfomance}

As the next step in assessing the basic cell performance under varying conditions, we investigate the transmission coefficients as a function of the environment temperature.
Qubit lifetimes are known to degrade with increasing temperature via two primary mechanisms\cite{Koch2007,Blais2021}:

(i) quasiparticle (QP) poisoning, which increases the relaxation rate $\Gamma_1$, and

(ii) elastic interactions with thermal photons, which do not exchange energy but induce random phase fluctuations of the qubit state, thereby increasing the dephasing rate $\Gamma_\varphi$.

We note that additional mechanisms (e.g., Andreev bound states) exist, but here we focus on these two dominant contributions for clarity.

In the basic cell, photon transfer between waveguides is strongly tied to the qubit’s coherence, as evidenced by flux-bias measurements where moving away from the sweet spot substantially reduced efficiency.
Consequently, one expects that fluctuations in the thermal photon number play the primary role in limiting the cell’s operation.

This expectation is further supported by the qubit’s coupling to two open waveguides with continuous spectra. Each waveguide carries thermal photons across a wide frequency range.
Each photon can be viewed as inducing a small, thermally driven "dispersive shift", effectively causing a random phase rotation of the qubit state via fluctuations in its transition frequency.
The cumulative effect of many such photons amplifies dephasing, and results the observed temperature-dependent reduction in coherence.

To examine this expectation quantitatively, we study the efficiency $E$ (Eq. \ref{eq:transfer_efficiency_with_dephasing}) as a function of the environment temperature by phenomenologically including both relaxation to the thermal bath $\Gamma_{bath}$ and dephasing $\Gamma_\varphi$ in the qubit frequency:

\begin{equation}
\label{eq:new_qb_freq_with_relax_deph}
\omega_{ge} \rightarrow \omega_{ge} - i \left(\Gamma_\varphi + \frac{\Gamma_{bath}}{2}\right).
\end{equation}
In contrast to the case of flux-bias dependence discussed in Section~\ref{sec:dephasing_due_to_flux},  we also account here for relaxation processes. This is because temperature variations can lead to energy loss through interactions with thermally activated quasiparticles, in addition to the dephasing effects.

We adopt simplified models for the temperature dependence of the relaxation and dephasing rates in terms of the thermal photon number $n_{th}$, neglecting spectral noise densities or more complex environmental features. We assume the commonly used linear dependencies:

\begin{align}
\Gamma_{bath} &= (2n_{th}+1)\,\gamma_1^0, \\
\Gamma_\varphi &= n_{th}\,\gamma_\varphi^0,
\end{align}
where $\gamma_1^0$ and $\gamma_\varphi^0$ are the zero-temperature relaxation and dephasing rates and can be interpreted as the rate increase per single thermal photon. For a more detailed discussion of models, we refer the reader to \cite{Koch2007,Yan2018,Lvov2025,Bertet2005}.

At finite temperature $T$, the thermal photon number can be estimated via Bose-Einstein statistics:

\begin{equation}
\label{eq:Bose-Einstein_stat}
n_{th} = \frac{1}{e^{\hbar\omega_{ge} / k_bT} - 1},
\end{equation}
where $k_b$ is the Boltzmann constant.
We note that this estimate considers only a single frequency component; a more complete treatment would integrate over the spectral density of the environment.
Combining these considerations, namely substituting Eqs. \ref{eq:new_qb_freq_with_relax_deph}-\ref{eq:Bose-Einstein_stat} into Eq.\ref{eq:our_efficiency_def} and considering the resonant case, the efficiency can be expressed as:

\begin{equation}
\label{eq:efficient_vs_thermal_phot}
E(n_{th}) = \frac{1}{1 + \left[n_{th}(\gamma_1^0 + \gamma_\varphi^0) + \frac{\gamma_1^0}{2}\right]\left(\frac{1}{\Gamma_A} + \frac{1}{\Gamma_B}\right) + \frac{\left[n_{th}(\gamma_1^0 + \gamma_\varphi^0) + \frac{\gamma_1^0}{2}\right]^2}{\Gamma_A \Gamma_B}}.
\end{equation}

\begin{figure}
    \centering
    \includegraphics[width=1\linewidth]{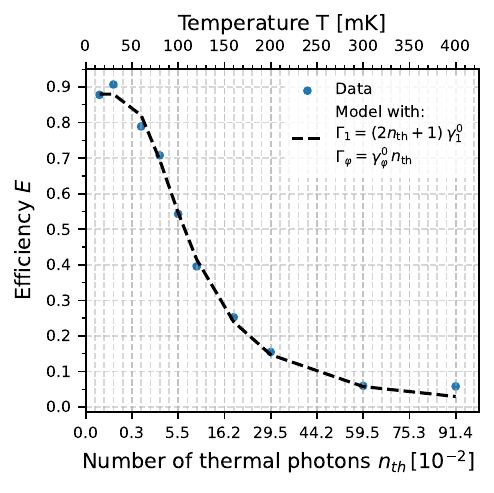}
    \caption{Temperature dependence of the basic cell efficiency as a function of the temperature.}
    \label{fig:temperature_dependence}
\end{figure}

The curve in Fig. \ref{fig:temperature_dependence} is obtained using Eq. \ref{eq:efficient_vs_thermal_phot}, with $\gamma_1^0 = 2\pi \cdot 0.26$~MHz, $\gamma_\varphi^0 = 2\pi \cdot 10.38$~MHz, and fixed values of $\Gamma_A$ and $\Gamma_B$ taken from the basic characterization chapter ($2\pi \cdot 1.81$~MHz and $2\pi \cdot 2.32$~MHz, respectively).

Since $\gamma_\varphi^0 \gg \gamma_1^0$, this confirms that dephasing dominates the basic cell performance, consistent with expectations.

\subsection{Photon number dependence and saturation}

In this section we study how the transmission coefficients of the basic cell depend on the input pulse power. For these measurements, the qubit is biased at its sweet spot. A 2~$\mu$s pulse at the qubit transition frequency $\omega_{ge}$ is applied, while its amplitude $A$ is varied. The corresponding pulse power is  $P=\frac{A^2}{2Z}$, where $Z=50$~Ohm is the impedance of the waveguides. The average photon number, thus, is $\langle n\rangle=\frac{P\cdot 2\mu s}{\hbar\omega_{ge}} $. For a single qubit in an open waveguide, the transmission magnitude is expected to scale approximately as  $\sim\frac{1}{\Omega^2/\Gamma^2} $, where $\Omega\sim \Gamma_{A/B}\sqrt{\langle n\rangle}$ and $\Gamma$ are drive and loss rates correspondingly \cite{Brehm2021,Astafiev2010}.We anticipate a similar qualitative behavior for the basic cell and adopt the following phenomenological fit function with fitting parameters $a,b,c,d$:
\begin{equation}
\label{eq:photon_number_expected_dependence}
    f(\langle n\rangle)=a-\frac{b}{1+\frac{\langle n\rangle^c}{d}}
\end{equation}

\begin{figure}[ht]
    \centering
    \includegraphics[width=1\linewidth]{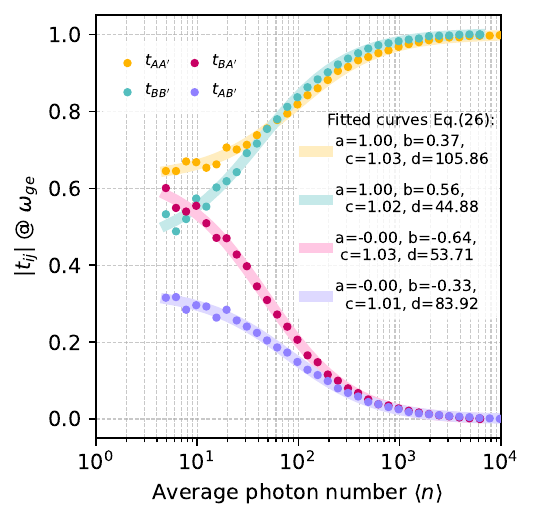}
    \caption{The basic cell performance as function of photon number in the probing signal. The calibrated transmission magnitudes, following the introduced procedure in the above sections, are shown along with the modeled curves Eq.(\ref{eq:photon_number_expected_dependence}).}
    \label{fig:photon_number_dependence}
\end{figure}

Figure~\ref{fig:photon_number_dependence} shows the magnitude of the transmission coefficients as a function of the average photon number, comparing calibrated responses $t_{ij}$. The extracted power exponent $c \approx 1$ confirms the expected linear photon number dependence ($\Omega^2\sim n$) and  that the basic cell operates as expected for $\langle n \rangle \lesssim 10$ photons. At high photon fluxes, the through coefficients approach unity, indicating that the qubit response is effectively suppressed under strong driving. Simultaneously, the cross coefficients drop toward zero, signaling transfer saturation: the qubit cannot process more than approximately one photon per cycle. This behavior can be interpreted as follows: in the cross coefficients $t_{AB}'$ and $t_{BA}'$, the incident field is naturally attenuated by the cell’s structure, whereas in the through coefficients we mostly detect photons that bypass the qubit without interaction. This observation suggests that the cross coefficients are more sensitive to qubit-mediated dynamics. Consequently, the basic cell inherently filters out unwanted control artifacts in the cross channels while preserving signals of physical relevance. Such a feature could be advantageous for probing other quantum phenomena, as explored in the following section.

\subsection{Platform for further research}

In the previous section, we made an assumption, that the cross coefficients are more sensitive to the qubit-mediated dynamics. Here we demonstrate that it is, in fact, a valuable feature -positioning the cell as a promising platform for exploring fundamental physical phenomena. To support this perspective, we present an additional measurement that qualitatively explains the observed nonlinearity using a straightforward, intuitive physical argument, without invoking a complex theoretical framework. This approach allows us to connect the observed results with more conventional measurement results, reinforcing the idea that the basic cell offers genuinely new opportunities for investigation.

When an electromagnetic field is applied to the qubit, the natural way to think about the processes that happens, is to consider the dressed state picture. Thus, the power of the field defines the number of photons $N$ at the frequency of the field $\omega$. These photons interact with the qubit, causing a qubit level splitting. This dressed-state structure can manifest in different observables: as Autler-Townes splitting in a probe-based spectroscopy measurement, or as the Mollow triplet in the resonance-fluorescence emission spectrum of the driven qubit. The splitting value could be defined in the simplified form (assuming single mode field) as \cite{Greenberg2017}:
\begin{equation*}
    \Omega_D=\sqrt{\left(\omega-\omega_{ge}\right)^2+4\Lambda^2N}
\end{equation*}
where $\Lambda$ is the coupling strength between the field and qubit. It means that the dressed transitions of the qubit are given by $\omega_{ge}\pm \frac{\Omega_D}{2}$. 

\begin{figure}[ht]
    \centering
    \includegraphics[width=\linewidth]{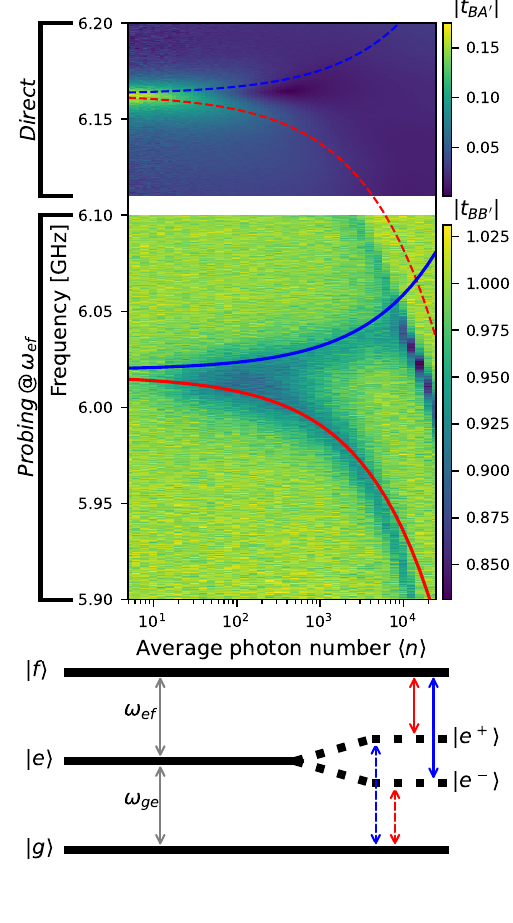}
    \caption{Dressing of first excited state $|e \rangle$ level observed directly via field detection in the cross-waveguide and indirectly by probing the subsequent transition between $|e \rangle$ and $|f \rangle$. The solid red and blue lines represent the extracted resonance frequencies from the $ |e\rangle \leftrightarrow |f\rangle$ probing experiment. The dashed red and blue lines are frequency-shifted copies of the solid lines, corresponding to the dressed-state transitions originating from the ground and excited manifolds, respectively.
The bottom panel shows a schematic diagram of the observed transitions.}
    \label{fig:dressing_magic}
\end{figure}

If the qubit becomes dressed by the field, it is natural to expect that the field itself may also be affected through backaction. In standard circuit QED setups, such dressing effects are typically observed at high field powers. However, the backaction of the qubit on the field is generally weak (quantum natured)  and masked by the strong driving field, making it difficult to detect directly. In these cases, one typically measures the radiation spectrum to observe peaks corresponding to dressed-state transitions, or introduces a weak probe signal to sense the level structure without significantly perturbing it. In contrast, at our basic cell, the input dressing field is suppressed when measured in the cross configuration due to isolation. This effectively creates a self-sustained filter, allowing detection of the qubit's backaction on the field, because the cross-channel contains primarily photons that have interacted with the qubit. To support this claim, we perform two experiments:

\begin{enumerate}
    \item In the first experiment, we send a signal with variable power and frequency near the qubit’s transition frequency $\omega_{ge}$. This dressing signal is long enough  (1~$\mu s$) to assume that the system reaches a steady-state response. This experiment can be viewed as a generalization of the previous results shown in Fig.~\ref{fig:photon_number_dependence}, where the signal was applied exactly on resonance with the qubit.
    Here, we send only a single signal to the waveguide \textit{B} and detect it in the waveguide \textit{$A^\prime$}. Explicitly, we measure the transmission coefficient \( t_{BA'}^{meas} \), where we intentionally omit the calibration procedure to demonstrate that the result can be achieved directly. As we use the same signal to dress the qubit and to detect the change of signal, we refer to this experiment as \textit{Direct}.
    \item  In the second experiment, we fix the   frequency of the dressing signal $\omega\approx\omega_{ge}$ and introduce additionally a weak probe signal near the \(\omega_{ef}\) transition, following the method described in \cite{Brehm2021}. The signal is detected it in the same waveguide output $B^\prime$. It means that our readout signal corresponds to $t_{BB^\prime}(\omega_{ef})$. This allows us to reconstruct the splitting \(\Omega_D\) for each dressing field power. We refer to this experiment as \textit{Probing}.
    
\end{enumerate}
The experimental results for both of these experiments are shown in Fig.\ref{fig:dressing_magic}.

From the \textit{Probing} experiment, we fit both shifted resonances as functions of photon number (shown as solid blue and red lines) using the following expressions:
\begin{align*}
    \omega_{ef}^{\text{red}} &= \omega_{ef} - \frac{1}{2} \sqrt{\left(\omega - \omega_{ge}\right)^2 + 4\Lambda_{\text{red}}^2 N}, \\
    \omega_{ef}^{\text{blue}} &= \omega_{ef} + \frac{1}{2} \sqrt{\left(\omega - \omega_{ge}\right)^2 + 4\Lambda_{\text{blue}}^2 N},
\end{align*}
with fitted coupling strengths $\Lambda_{\text{red}} = 2\pi \cdot 0.81~\text{MHz}$ and $\Lambda_{\text{blue}} = 2\pi \cdot 0.39~\text{MHz}$.

In the \textit{Direct} measurements, both $\omega$ and $N$ are varied simultaneously, resulting in shifts that would require a full nonlinear model to accurately capture them. Developing such a model is beyond the scope of this work and is left for a future study. Instead, for simplicity, we apply the same type of shift to the $\omega_{ge}$ transition, using:
\begin{align*}
    \omega_{ge}^{\text{red}} &= \omega_{ge} - \frac{1}{2} \sqrt{\left(\omega - \omega_{ge}\right)^2 + 4\Lambda_{\text{red}}^2 N}, \\
    \omega_{ge}^{\text{blue}} &= \omega_{ge} + \frac{1}{2} \sqrt{\left(\omega - \omega_{ge}\right)^2 + 4\Lambda_{\text{blue}}^2 N}.
\end{align*}
These are shown as dashed red and blue lines in Fig.~\ref{fig:dressing_magic}. Interestingly, the red dashed line qualitatively explains a third-line feature observed at high powers in the \textit{Probing} experiment going toward lower frequencies as the photon number increase, a distinct red-shifted line corresponding to the $\omega_{ge}$ transition. Since the dressing field in this experiment was resonant with the $e \leftrightarrow g$ transition, the observed shift is larger than what the red dashed line predicts, which is consistent with stronger dressing in this case.

Although we do not yet provide a complete quantitative model, the correspondence between our simplified analysis and the experimental features strongly supports the interpretation that the basic cell enables direct observation of qubit dressing through backaction on the driving field, measured in the cross-transmission configuration. Remarkably, this effect is visible without relying on conventional probe spectroscopy or spectral reconstruction, highlighting the cell's ability to act as a self-isolated platform where the field-qubit interaction manifests in the field itself.

This demonstrates that the basic cell is not only a robust and scalable component, but also a uniquely capable platform for fundamental studies of light–matter interaction in the strong-driving regime. Its ability to isolate and expose backaction effects, typically masked in standard setups, opens new avenues for exploring nonlinearity, dressed states, and other emergent quantum phenomena under minimal experimental overhead. As such, the cell represents a promising tool for both practical applications and foundational research in circuit quantum electrodynamics.

\section{Conclusion}
We have introduced and implemented a novel basic cell for a quantum router, accompanied by a theoretical framework capturing its dynamics and interactions. Experimentally, we validated the cell’s operation, extracted its key parameters, and demonstrated consistency with theoretical predictions. Complementary studies — including flux-bias, photon-number, and temperature dependence — allowed us to probe operational limits and quantify the robustness of the system. In particular, the flux and temperature-dependent measurements revealed that dephasing is the dominant mechanism limiting performance, in agreement with theoretical expectations. Importantly, beyond its primary routing function, the basic cell also provides a versatile platform for fundamental investigations of quantum coherence and photon-mediated interactions, opening avenues for further experimental studies and exploration of complex quantum phenomena.

\begin{acknowledgments}
This work was partially supported by the German Federal Ministry of Education and Research under Grant Nos. 13N16152/QSolid, 13N16258/SuperLSI and the European Innovation Council’s Pathfinder Open programme under grant agreement number 101129663/QRC-4-ESP.
\end{acknowledgments}
\section*{Authors contributions}

% \section*{Data Availability Statement}

% AIP Publishing believes that all datasets underlying the conclusions of the paper should be available to readers. Authors are encouraged to deposit their datasets in publicly available repositories or present them in the main manuscript. All research articles must include a data availability statement stating where the data can be found. In this section, authors should add the respective statement from the chart below based on the availability of data in their paper.

\appendix

\section{Effective S-matrix derivation }
\label{app:Seffective}
We start from the standard definition of an S-matrix for any circuit element:
\begin{equation} \label{eq:smat_definition_new}
S_i \vec{a_i} = \vec{b_i},
\end{equation}
where $i$ labels a specific circuit element (see Fig.\ref{fig:matrix_scheme}), $\vec{a_i}$ is the vector of incoming waves, and $\vec{b_i}$ is the vector of outgoing waves. For input/output lines, $S_i$ is a $2\times2$ matrix; for the basic cell, $S_i$ is $4\times4$.

For clarity in deriving the effective matrix, we separate waves into \textit{external} (connected to measurement lines) and \textit{internal} (connected to the basic cell). We define:
\begin{align}\label{eq:grouping_order_new}
\vec{a}_{ext} &= (a_{1A},a_{2GA},a_{1B},a_{2GB}), &
\vec{a}_{int} &= (a_{2A},a_{1GA},a_{2B},a_{1GB}) \\
\vec{b}_{ext} &= (b_{1A},b_{2GA},b_{1B},b_{2GB}), &
\vec{b}_{int} &= (b_{2A},b_{1GA},b_{2B},b_{1GB}).
\end{align}

Physically, this grouping distinguishes waves we can measure (\textit{external}) from those that circulate inside the device (\textit{internal}). This separation allows us to systematically eliminate internal degrees of freedom to obtain the effective S-matrix.

We note, that internal waves also could be defined in terms of waves incoming and outcoming to the basic cell:
\begin{align*}
    \vec{a}_{int}=(b_1,b_2,b_3,b_4) \\
    \vec{b}_{int}=(a_1,a_2,a_3,a_4)
\end{align*}
where a swap between notations $a$ and $b$ is done, as the same wave could be an incoming for external circuity element and an outcoming from the basic cell. It means, that the S-matrix of the basic cell could be also defined in the same grouped vectors as $S\vec{b}_{int}= \vec{a}_{int}$.

Our goal is to derive a $4\times4$ effective S-matrix that directly relates $\vec{a}_{ext}$ to $\vec{b}_{ext}$. To this end, we define a complementary matrix $S_{comp}$ that accounts for all internal and external connections:
\begin{equation}\label{eq:supporting_mat_new}
S_{comp} \begin{pmatrix}
\vec{a}_{ext} \\ \vec{a}_{int}
\end{pmatrix}
= \begin{pmatrix}
\vec{b}_{ext} \\ \vec{b}_{int} 
\end{pmatrix}.
\end{equation}

This matrix allows us to systematically eliminate internal waves later, leading to the effective S-matrix for the externally measurable quantities. Conceptually, $S_{comp}$ encodes the full response of the circuit, including both measured and hidden paths.

The natural block-diagonal representation of the circuit is
\begin{equation}\label{eq:natural_order_full}
\begin{pmatrix}
S_A & 0 & 0 & 0 \\
0 & G_A & 0 & 0 \\
0 & 0 & S_B & 0 \\
0 & 0 & 0 & S_B
\end{pmatrix}
\begin{pmatrix} \vec{a}_A \\ \vec{a}_{GA} \\ \vec{a}_B \\ \vec{a}_{GB} \end{pmatrix}
=
\begin{pmatrix} \vec{b}_A \\ \vec{b}_{GA} \\ \vec{b}_B \\ \vec{b}_{GB} \end{pmatrix}.
\end{equation}
Please note that here $0$ is 2-by-2 matrix of zeros.
\noindent To match the previously defined external/internal grouping, we introduce a permutation matrix $P$:
\begin{equation*}
P\begin{pmatrix} \vec{a}_A \\ \vec{a}_{GA} \\ \vec{a}_B \\ \vec{a}_{GB} \end{pmatrix}
= \begin{pmatrix} \vec{a}_{ext} \\ \vec{a}_{int} \end{pmatrix}, \quad
P\begin{pmatrix} \vec{b}_A \\ \vec{b}_{GA} \\ \vec{b}_B \\ \vec{b}_{GB} \end{pmatrix}
= \begin{pmatrix} \vec{b}_{ext} \\ \vec{b}_{int} \end{pmatrix}.
\end{equation*}

This step ensures consistency in notation and prepares the system for block elimination. Physically, it is simply a reordering of indices to separate measurable and internal waves. The explicit form of the permutation matrix is:
\begin{equation*}
P=\begin{pmatrix}
    1 & 0 & 0 & 0 & 0 & 0 & 0 & 0 \\
    0 & 0 & 0 & 1 & 0 & 0 & 0 & 0 \\
    0 & 0 & 0 & 0 & 1 & 0 & 0 & 0 \\
    0 & 0 & 0 & 0 & 0 & 0 & 0 & 1 \\
    0 & 1 & 0 & 0 & 0 & 0 & 0 & 0 \\
    0 & 0 & 1 & 0 & 0 & 0 & 0 & 0 \\
    0 & 0 & 0 & 0 & 0 & 1 & 0 & 0 \\
    0 & 0 & 0 & 0 & 0 & 0 & 1 & 0 \\
\end{pmatrix}.
\end{equation*}

It allows us to rewrite equation (\ref{eq:natural_order_full}) via several steps as: 
\begin{align*}
        \begin{pmatrix}
        S_A & 0 & 0 & 0 \\
        0 & G_A & 0 & 0 \\
        0 & 0 & S_B & 0 \\
        0 & 0 & 0 & S_B \\
    \end{pmatrix}   P^{-1}\begin{pmatrix}
    \vec{a}_{ext}\\
    \vec{a}_{int}
\end{pmatrix} = P^{-1}\begin{pmatrix}
    \vec{b}_{ext}\\
    \vec{b}_{int}
\end{pmatrix}\\
P\begin{pmatrix}
        S_A & 0 & 0 & 0 \\
        0 & G_A & 0 & 0 \\
        0 & 0 & S_B & 0 \\
        0 & 0 & 0 & S_B \\
    \end{pmatrix}P^{-1} \begin{pmatrix}
    \vec{a}_{ext}\\
    \vec{a}_{int}
\end{pmatrix} =\begin{pmatrix}
    \vec{b}_{ext}\\
    \vec{b}_{int}
\end{pmatrix}\\
\end{align*}

which is the same form as complementary matrix equation \ref{eq:supporting_mat_new}.
It could be shown, that for the chosen permutation matrix the compact form of this equation is:
 \begin{align}
    S_{11} \vec{a}_{ext}+S_{12}\vec{a}_{int}=\vec{b}_{ext}\\
    S_{21} \vec{a}_{ext}+S_{22}\vec{a}_{int}=\vec{b}_{int}
\end{align}
with 
\begin{align*}
S_{11}^{comp}=\text{diag}(S_{11}^{(A)}, S_{11}^{(Ga)}, S_{11}^{(B)}, S_{11}^{(Gb)}), \\
S_{12}^{comp}=\text{diag}(S_{12}^{(A)}, S_{12}^{(Ga)}, S_{12}^{(B)}, S_{12}^{(Gb)}) \\
S_{21}^{comp}=\text{diag}(S_{21}^{(A)}, S_{21}^{(Ga)}, S_{21}^{(B)}, S_{21}^{(Gb)}) \\
S_{22}^{comp}=\text{diag}(S_{22}^{(A)}, S_{22}^{(Ga)}, S_{22}^{(B)}, S_{22}^{(Gb)}) \\
\end{align*}

where $S_{nm}^{i}$  are element at $nm$-position of the S matrix of element $i$ and $diag$ stands for diagonal matrix with corresponding elements.
Eliminating the internal waves using
\begin{equation}
\vec{b}_{int} = S^{-1} \vec{a}_{int},
\end{equation}
we obtain the effective matrix relating only external waves:
\begin{equation}
S^{meas} = S_{11} + S_{12} (S^{-1} - S_{22})^{-1} S_{21}, \quad
S^{meas}\vec{a}_{ext} = \vec{b}_{ext}.
\end{equation}

This is the S-matrix that describes what we actually measure. In the limit of negligible reflections ($S_{nn}^{i}=0$) and unit transmission ($S_{nm}^{i}=1$), $S^{meas}$ reduces to the basic cell matrix, confirming the consistency of the derivation.

To account for multiple reflections inside the circuit, we can expand
\begin{equation}
S^{meas} = S_{11} + S_{12} S \sum_{k=0}^{\infty} (S S_{22})^k S_{21},
\end{equation}
using the Neumann series $(I - X)^{-1} = \sum_{k} X^k$.

Physically, each term $k$ represents waves that undergo $k$ internal reflections before reaching the external ports. For weak reflections, higher-order terms rapidly decay, and we can approximate
\begin{equation}
S^{meas} \approx S_{11} + S_{12} S S_{21},
\end{equation}
which justifies the simplified expressions used in the main text for $t_{AA^\prime}, t_{AB^\prime}, t_{BA^\prime}, t_{BB^\prime}$.
We note, that we did not explicitly include isolation in this full block-matrix derivation. Instead, we introduced an additive isolation term to the cross-transmission coefficients. This approximation is justified as the isolation term is small, and obtained agreement between theory and experiment confirms that its contribution does not significantly affect the effective S-matrix.
\section{Time domain raw data}\label{app:raw_time_domain}

Here we present raw data of time-domain experiments for reference, before we applied the decomposition procedure mentioned in the main text. They demonstrate the same difference in the DC offsets as at Fig. \ref{fig:raw_rabi_amplitudes} of the main text.  Moreover, there raw readout amplitudes corresponding to the qubit in different states are matching with ones observed in $\pi$-amplitude calibration results.  
\begin{widetext}
    
\begin{figure*}[!hb]
    \centering
    \includegraphics[width=1\textwidth]{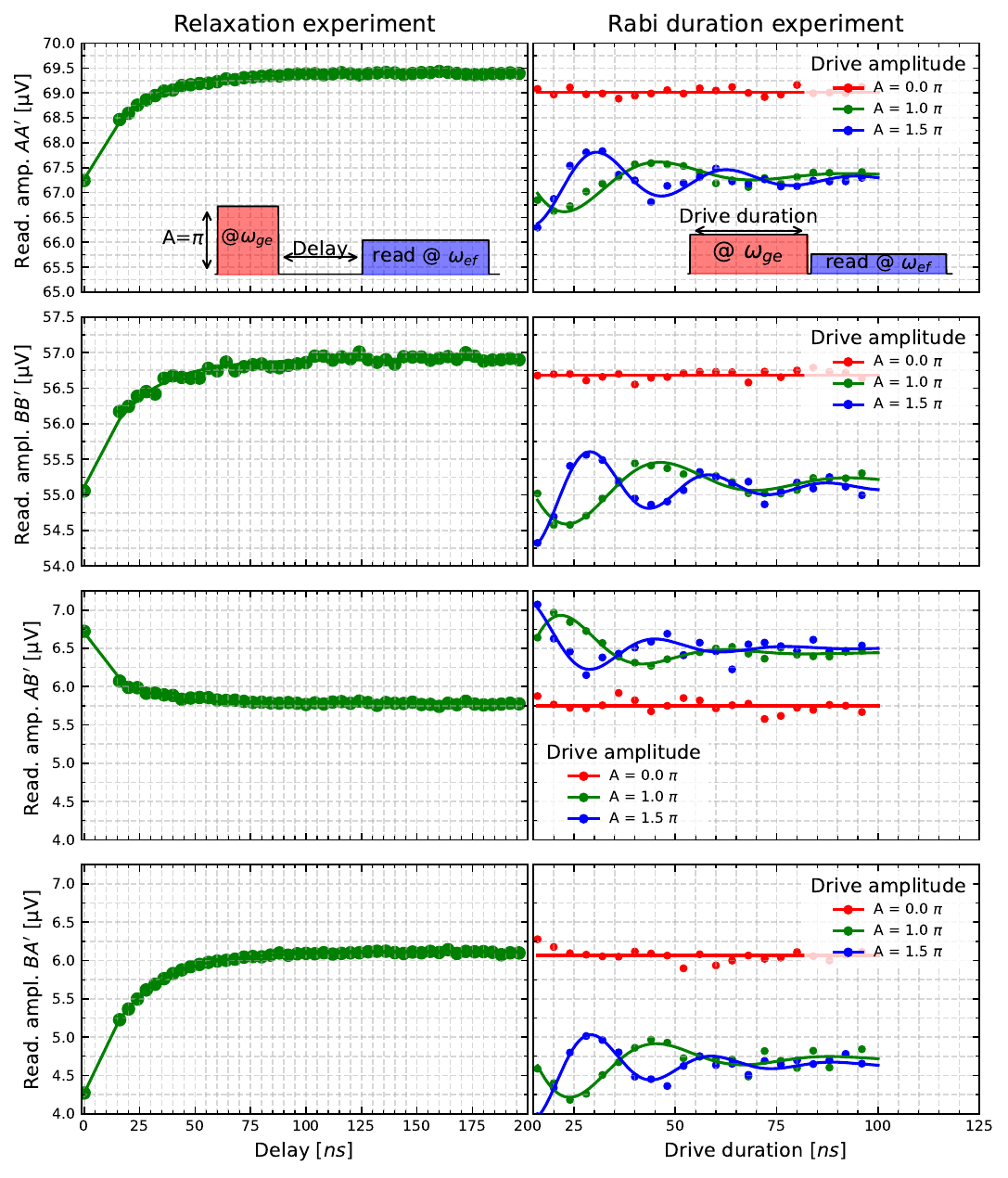}
    \caption{Rabi oscillation experiment with varying drive pulse amplitude. The drive pulse is sent to one of two waveguides at 01-transition frequency, the readout pulse is sent to the same waveguide as the drive pulse, but detected either there (through) or in opposite waveguide's output port (cross). The data is shown in raw format, obtained with a standard heterodyne detection scheme.}
    \label{fig:raw_rabi_lifetimes}
\end{figure*}
\end{widetext}

\clearpage

\nocite{*}
\bibliography{aipsamp}

@article{Koch2007,
 author = {Koch, Jens and Yu, Terri M. and Gambetta, Jay and Houck, A. A. and Schuster, D. I. and Majer, J. and Blais, Alexandre and Devoret, M. H. and Girvin, S. M. and Schoelkopf, R. J.},
title = {Charge-insensitive qubit design derived from the Cooper pair box},
  journal = {Phys. Rev. A},
  volume = {76},
  issue = {4},
  pages = {042319},
  numpages = {19},
  year = {2007},
  month = {Oct},
  publisher = {American Physical Society},
  doi = {10.1103/PhysRevA.76.042319},
 
}

@article{Brehm2021,
  author       = {Jan David Brehm and Alexander N. Poddubny and Alexander Stehli and Tim Wolz and Hannes Rotzinger and Alexey V. Ustinov},
  title        = {Waveguide bandgap engineering with an array of superconducting qubits},
  journal      = {npj Quantum Materials},
  year         = {2021},
  volume       = {6},
  number       = {1},
  pages        = {10},
  doi          = {10.1038/s41535-021-00310-z},
  url          = {https://doi.org/10.1038/s41535-021-00310-z},
  issn         = {2397-4648},
}

@article{Greenberg2017,
  title = {Mollow triplet through pump-probe single-photon spectroscopy of artificial atoms},
  author = {Greenberg, Ya. S. and Sultanov, A. N.},
  journal = {Phys. Rev. A},
  volume = {95},
  issue = {5},
  pages = {053840},
  numpages = {16},
  year = {2017},
  month = {May},
  publisher = {American Physical Society},
  doi = {10.1103/PhysRevA.95.053840},
  url = {https://link.aps.org/doi/10.1103/PhysRevA.95.053840}
}

@article{Probst2015,
    author = {Probst, S. and Song, F. B. and Bushev, P. A. and Ustinov, A. V. and Weides, M.},
    title = {Efficient and robust analysis of complex scattering data under noise in microwave resonators},
    journal = {Review of Scientific Instruments},
    volume = {86},
    number = {2},
    pages = {024706},
    year = {2015},
    month = {02},
    issn = {0034-6748},
    doi = {10.1063/1.4907935},
    url = {https://doi.org/10.1063/1.4907935}
}

@article{Greenberg2015,
  author = {Greenberg, Ya. S. and Shtygashev, A. A.},
title = {Non-Hermitian Hamiltonian approach to the microwave transmission through a one-dimensional qubit chain},
  journal = {Phys. Rev. A},
  volume = {92},
  issue = {6},
  pages = {063835},
  numpages = {15},
  year = {2015},
  month = {Dec},
  doi = {10.1103/PhysRevA.92.063835}}

@Article{Sultanov2020,
AUTHOR = {Sultanov, Aydar and Greenberg, Yakov and Mutsenik, Evgeniya and Pitsun, Dmitry and Il’ichev, Evgeni},
title = {Universal Tool for Single-Photon Circuits: Quantum Router Design},
JOURNAL = {Materials},
VOLUME = {13},
YEAR = {2020},
NUMBER = {2},
ARTICLE-NUMBER = {319},
PubMedID = {32284507},
ISSN = {1996-1944},
DOI = {10.3390/ma13020319}
}

@ARTICLE{Schmelz2024,
  author={Schmelz, Matthias and Mutsenik, E. and Bravin, S. and Sultanov, A. and Ziegler, M. and Hübner, U. and Peiselt, K. and Mechold, S. and Oelsner, G. and Kunert, J. and Stolz, R.},
  journal={IEEE Transactions on Applied Superconductivity}, 
  title={Wafer-Scale Al Junction Technology for Superconducting Quantum Circuits}, 
  year={2024},
  volume={34},
  number={3},
  pages={1-5},
  doi={10.1109/TASC.2024.3350580}}

@ARTICLE{Kaczmarek2025,
  author={Kaczmarek, L. and Schmelz, M. and Peiselt, K. and Sultanov, A. and Mutsenik, E. and Oelsner, G. and Stolz, R.},
  journal={IEEE Transactions on Applied Superconductivity}, 
  title={Wafer-Scale Fabrication Technologies for Integrated Superconducting Quantum Circuits}, 
  year={2025},
  volume={35},
  number={5},
  pages={1-4},
  doi={10.1109/TASC.2024.3516747}}

@article{Astafiev2010,
author = {O. Astafiev  and A. M. Zagoskin  and A. A. Abdumalikov  and Yu. A. Pashkin  and T. Yamamoto  and K. Inomata  and Y. Nakamura  and J. S. Tsai },
title = {Resonance Fluorescence of a Single Artificial Atom},
journal = {Science},
volume = {327},
number = {5967},
pages = {840-843},
year = {2010},
doi = {10.1126/science.1181918}}

@article{Sultanov2025,
    author = {Sultanov, A. and Mutsenik, E. and Schmelz, M. and Kaczmarek, L. and Oelsner, G. and Hübner, U. and Stolz, R. and Il'ichev, E.},
    title = {Measuring coherent dynamics of a superconducting qubit in an open waveguide},
    journal = {Applied Physics Letters},
    volume = {127},
    number = {4},
    pages = {042602},
    year = {2025},
    month = {07},
    issn = {0003-6951},
    doi = {10.1063/5.0277075},
    url = {https://doi.org/10.1063/5.0277075},
}

@article{Kimble2008,
author = { H. I. Kimble   },
title = {The quantum internet},
journal = {Nature},
volume = {453},
number = {},
pages = {1023–1030},
year = {2008},
doi = { doi:10.1038/nature07127}
}

@article{Walther2006,
author = {Herbert Walther, Benjamin T H Varcoe, Berthold-Georg Englert and Thomas Becker},
title = {Cavity quantum electrodynamics},
journal = {Rep. Prog. Phys.},
volume = {69},
number = {},
pages = {1325},
year = {2006},
doi = { doi:10.1088/0034-4885/69/5/R02}
}

@article{Blais2021,
  title = {Circuit quantum electrodynamics},
  author = {Blais, Alexandre and Grimsmo, Arne L. and Girvin, S. M. and Wallraff, Andreas},
  journal = {Rev. Mod. Phys.},
  volume = {93},
  issue = {2},
  pages = {025005},
  numpages = {72},
  year = {2021},
  month = {May},
  publisher = {American Physical Society},
  doi = {10.1103/RevModPhys.93.025005},
  url = {https://link.aps.org/doi/10.1103/RevModPhys.93.025005}
}

@article{Sheremet2023,
  title = {Waveguide quantum electrodynamics: Collective radiance and photon-photon correlations},
  author = {Sheremet, Alexandra S. and Petrov, Mihail I. and Iorsh, Ivan V. and Poshakinskiy, Alexander V. and Poddubny, Alexander N.},
  journal = {Rev. Mod. Phys.},
  volume = {95},
  issue = {1},
  pages = {015002},
  numpages = {59},
  year = {2023},
  month = {Mar},
  publisher = {American Physical Society},
  doi = {10.1103/RevModPhys.95.015002},
  url = {https://link.aps.org/doi/10.1103/RevModPhys.95.015002}
}

@article{Delsing2011,
  title = {Demonstration of a Single-Photon Router in the Microwave Regime},
  author = {Hoi, Io-Chun and Wilson, C. M. and Johansson, G\"oran and Palomaki, Tauno and Peropadre, Borja and Delsing, Per},
  journal = {Phys. Rev. Lett.},
  volume = {107},
  issue = {7},
  pages = {073601},
  numpages = {5},
  year = {2011},
  month = {Aug},
  publisher = {American Physical Society},
  doi = {10.1103/PhysRevLett.107.073601},
  url = {https://link.aps.org/doi/10.1103/PhysRevLett.107.073601}
}

@article{Yan2014,
  title = {Single-photon quantum router with multiple output ports},
  author = {Yan, Wei-Bin and Fan, Heng },
  journal = {Scientific Reports},
  volume = {4},
  issue = {},
  pages = {4820},
  numpages = {},
  year = {2014},
  month = {Apr},
  publisher = {},
  doi = {10.1038/srep04820},
  url = {}
}

@article{Zhu2019,
  title = {Single-photon quantum router in the microwave regime utilizing double superconducting
resonators with tunable coupling},
  author = {Zhu, Y. T. and Jia, C. M.},
  journal = {Phys. Rev. A.},
  volume = {99},
  issue = {},
  pages = {063815},
  numpages = {},
  year = {2019},
  month = {Jun},
  publisher = {American Physical Society},
  doi = {10.1103/PhysRevA.99.063815},
  url = {}
}

@article{Gong2024,
  title = {Tunable quantum router with giant atoms, implementing quantum gates,
teleportation, non-reciprocity, and circulators},
  author = {Gong, Rui-Yang and He, Zi-Yu and Yu, Cheng-He and Zhang, Ge-Fei and Nori, Franko and  Xiang,  Ze-Liang},
  journal = {arXiv:2411.19307v1 [quant-ph]},
  volume = {},
  issue = {},
  pages = {},
  numpages = {},
  year = {2024},
  month = {Nov},
  publisher = {},
  doi = {},
  url = {}
}

@article{Bartholomew2020,
  title = {On-chip coherent microwave-to-optical transduction mediated by ytterbium in YVO4},
  author = {Bartholomew, John G. and Rochman, Jake and Xie, Tian and  Kindem, Jonathan M. and Ruskuc, Andrei and  Craiciu,  Ioana and Lei, Mi and Faraon, Andrei},
  journal = {Nat. Communications},
  volume = {11},
  issue = {},
  pages = {3266},
  numpages = {},
  year = {2020},
  month = {Jun},
  publisher = {},
  doi = {10.1038/s41467-020-16996-x},
  url = {}
}

@article{Wang2018,
  title = {Integrated lithium niobate electro-optic modulators operating at CMOS-compatible voltages},
  author = {Wang, Cheng and Zhang, Mian and Chen, Xi and  Bertrand, Maxime and Shams-Ansari, Amirhassan and  Chandrasekhar,  Sethumadhavan and Winzer, Peter and Lončar, Marko},
  journal = {Nature},
  volume = {562},
  issue = {},
  pages = {101–104},
  numpages = {},
  year = {2018},
  month = {Sep},
  publisher = {},
  doi = {10.1038/s41586-018-0551-y},
  url = {}
}

@article{Wang2022,
  title = {Integrated lithium niobate electro-optic modulators operating at CMOS-compatible voltages},
  author = {Wang, Changqing and Gonin, Ivan and Grassellino, Anna and  Kazakov, Sergey and Romanenko, Alexander and  Yakovlev,  Vyacheslav P. and Zorzetti, Silvia},
  journal = {npj Quantum Inf},
  volume = {8},
  issue = {},
  pages = {149},
  numpages = {},
  year = {2022},
  month = {Des},
  publisher = {},
  doi = {10.1038/s41534-022-00664-7}}

@article{Ithier2005,
  title = {Decoherence in a superconducting quantum bit circuit},
  author = {Ithier, G. and Collin, E. and Joyez, P. and Meeson, P. J. and Vion, D. and Esteve, D. and Chiarello, F. and Shnirman, A. and Makhlin, Y. and Schriefl, J. and Sch\"on, G.},
  journal = {Phys. Rev. B},
  volume = {72},
  issue = {13},
  pages = {134519},
  numpages = {22},
  year = {2005},
  month = {Oct},
  publisher = {American Physical Society},
  doi = {10.1103/PhysRevB.72.134519},
  url = {https://link.aps.org/doi/10.1103/PhysRevB.72.134519}
}

@article{Lupascu2013,
    author = {Deng, Chunqing and Otto, Martin and Lupascu, Adrian},
    title = {An analysis method for transmission measurements of superconducting resonators with applications to quantum-regime dielectric-loss measurements},
    journal = {Journal of Applied Physics},
    volume = {114},
    number = {5},
    pages = {054504},
    year = {2013},
    month = {08},
    issn = {0021-8979},
    doi = {10.1063/1.4817512},
    url = {https://doi.org/10.1063/1.4817512}}

@article{Rieger2023,
  title = {Fano Interference in Microwave Resonator Measurements},
  author = {Rieger, D. and G\"unzler, S. and Spiecker, M. and Nambisan, A. and Wernsdorfer, W. and Pop, I.M.},
  journal = {Phys. Rev. Appl.},
  volume = {20},
  issue = {1},
  pages = {014059},
  numpages = {14},
  year = {2023},
  month = {Jul},
  publisher = {American Physical Society},
  doi = {10.1103/PhysRevApplied.20.014059},
  url = {https://link.aps.org/doi/10.1103/PhysRevApplied.20.014059}
}

@article{Biancetti2009,
  title = {Dynamics of dispersive single-qubit readout in circuit quantum electrodynamics},
  author = {Bianchetti, R. and Filipp, S. and Baur, M. and Fink, J. M. and G\"oppl, M. and Leek, P. J. and Steffen, L. and Blais, A. and Wallraff, A.},
  journal = {Phys. Rev. A},
  volume = {80},
  issue = {4},
  pages = {043840},
  numpages = {7},
  year = {2009},
  month = {Oct},
  publisher = {American Physical Society},
  doi = {10.1103/PhysRevA.80.043840},
  url = {https://link.aps.org/doi/10.1103/PhysRevA.80.043840}
}

@article{Biancetti2010,
  title = {Control and Tomography of a Three Level Superconducting Artificial Atom},
  author = {Bianchetti, R. and Filipp, S. and Baur, M. and Fink, J. M. and Lang, C. and Steffen, L. and Boissonneault, M. and Blais, A. and Wallraff, A.},
  journal = {Phys. Rev. Lett.},
  volume = {105},
  issue = {22},
  pages = {223601},
  numpages = {4},
  year = {2010},
  month = {Nov},
  publisher = {American Physical Society},
  doi = {10.1103/PhysRevLett.105.223601},
  url = {https://link.aps.org/doi/10.1103/PhysRevLett.105.223601}
}

@article{Yan2018,
  title = {Distinguishing Coherent and Thermal Photon Noise in a Circuit Quantum Electrodynamical System},
  author = {Yan, Fei and Campbell, Dan and Krantz, Philip and Kjaergaard, Morten and Kim, David and Yoder, Jonilyn L. and Hover, David and Sears, Adam and Kerman, Andrew J. and Orlando, Terry P. and Gustavsson, Simon and Oliver, William D.},
  journal = {Phys. Rev. Lett.},
  volume = {120},
  issue = {26},
  pages = {260504},
  numpages = {6},
  year = {2018},
  month = {Jun},
  publisher = {American Physical Society},
  doi = {10.1103/PhysRevLett.120.260504},
  url = {https://link.aps.org/doi/10.1103/PhysRevLett.120.260504}
}

@article{Lvov2025,
  title = {Thermometry based on a superconducting qubit},
  author = {Lvov, D. S. and Lemziakov, S. A. and Ankerhold, E. and Peltonen, J. T. and Pekola, J. P.},
  journal = {Phys. Rev. Appl.},
  volume = {23},
  issue = {5},
  pages = {054079},
  numpages = {20},
  year = {2025},
  month = {May},
  publisher = {American Physical Society},
  doi = {10.1103/PhysRevApplied.23.054079},
  url = {https://link.aps.org/doi/10.1103/PhysRevApplied.23.054079}
}

@article{Bertet2005,
  title = {Dephasing of a Superconducting Qubit Induced by Photon Noise},
  author = {Bertet, P. and Chiorescu, I. and Burkard, G. and Semba, K. and Harmans, C. J. P. M. and DiVincenzo, D. P. and Mooij, J. E.},
  journal = {Phys. Rev. Lett.},
  volume = {95},
  issue = {25},
  pages = {257002},
  numpages = {4},
  year = {2005},
  month = {Dec},
  publisher = {American Physical Society},
  doi = {10.1103/PhysRevLett.95.257002},
  url = {https://link.aps.org/doi/10.1103/PhysRevLett.95.257002}
}

@book{jolliffe2002pca,
  author    = {Ian T. Jolliffe},
  title     = {Principal Component Analysis},
  year      = {2002},
  edition   = {2},
  publisher = {Springer},
  address   = {New York},
  series    = {Springer Series in Statistics},
  isbn      = {978-0-387-95442-4},
  doi       = {10.1007/b98835}
}

@article{Azcona2025,
  title = {Quantum Dynamics with Stochastic Non-Hermitian Hamiltonians},
  author = {Martinez-Azcona, Pablo and Kundu, Aritra and Saxena, Avadh and del Campo, Adolfo and Chenu, Aur\'elia},
  journal = {Phys. Rev. Lett.},
  volume = {135},
  issue = {1},
  pages = {010402},
  numpages = {6},
  year = {2025},
  month = {Jul},
  publisher = {American Physical Society},
  doi = {10.1103/5ksl-tjjm},
  url = {https://link.aps.org/doi/10.1103/5ksl-tjjm}
}

@article{Molmer2013,
  title = {Effective description of tunneling in a time-dependent potential with applications to voltage switching in Josephson junctions},
  author = {Andersen, Christian Kraglund and M\o{}lmer, Klaus},
  journal = {Phys. Rev. A},
  volume = {87},
  issue = {5},
  pages = {052119},
  numpages = {7},
  year = {2013},
  month = {May},
  publisher = {American Physical Society},
  doi = {10.1103/PhysRevA.87.052119},
  url = {https://link.aps.org/doi/10.1103/PhysRevA.87.052119}
}

@article{Dalibard1992,
  title = {Wave-function approach to dissipative processes in quantum optics},
  author = {Dalibard, Jean and Castin, Yvan and M\o{}lmer, Klaus},
  journal = {Phys. Rev. Lett.},
  volume = {68},
  issue = {5},
  pages = {580--583},
  numpages = {0},
  year = {1992},
  month = {Feb},
  publisher = {American Physical Society},
  doi = {10.1103/PhysRevLett.68.580},
  url = {https://link.aps.org/doi/10.1103/PhysRevLett.68.580}
}

@article{Passante1993,
title = {Virtual transitions, self-dressing and indirect spectroscopy},
journal = {Optics Communications},
volume = {99},
number = {1},
pages = {55-60},
year = {1993},
issn = {0030-4018},
doi = {https://doi.org/10.1016/0030-4018(93)90704-9},
url = {https://www.sciencedirect.com/science/article/pii/0030401893907049},
author = {R. Passante and T. Petrosky and I. Prigogine}}

@article{Karpov2000,
    author = {Karpov, E. and Prigogine, I. and Petrosky, T. and Pronko, G.},
    title = {Friedrichs model with virtual transitions. Exact solution and indirect spectroscopy},
    journal = {Journal of Mathematical Physics},
    volume = {41},
    number = {1},
    pages = {118-131},
    year = {2000},
    month = {01},
    issn = {0022-2488},
    doi = {10.1063/1.533125},
    url = {https://doi.org/10.1063/1.533125},
    
}

@article{Liu2024,
  title = {Observation of Discrete Charge States of a Coherent Two-Level System in a Superconducting Qubit},
  author = {Liu, Bao-Jie and Wang, Ying-Ying and Sheffer, Tal and Wang, Chen},
  journal = {Phys. Rev. Lett.},
  volume = {133},
  issue = {16},
  pages = {160602},
  numpages = {7},
  year = {2024},
  month = {Oct},
  publisher = {American Physical Society},
  doi = {10.1103/PhysRevLett.133.160602},
  url = {https://link.aps.org/doi/10.1103/PhysRevLett.133.160602}
}

@article{Hutin2024,
  title = {Monitoring the Energy of a Cavity by Observing the Emission of a Repeatedly Excited Qubit},
  author = {Hutin, H. and Essig, A. and Assouly, R. and Rouchon, P. and Bienfait, A. and Huard, B.},
  journal = {Phys. Rev. Lett.},
  volume = {133},
  issue = {15},
  pages = {153602},
  numpages = {6},
  year = {2024},
  month = {Oct},
  publisher = {American Physical Society},
  doi = {10.1103/PhysRevLett.133.153602},
  url = {https://link.aps.org/doi/10.1103/PhysRevLett.133.153602}
}

@article{Hamann2018,
  title = {Nonreciprocity Realized with Quantum Nonlinearity},
  author = {Rosario Hamann, Andr\'es and M\"uller, Clemens and Jerger, Markus and Zanner, Maximilian and Combes, Joshua and Pletyukhov, Mikhail and Weides, Martin and Stace, Thomas M. and Fedorov, Arkady},
  journal = {Phys. Rev. Lett.},
  volume = {121},
  issue = {12},
  pages = {123601},
  numpages = {5},
  year = {2018},
  month = {Sep},
  publisher = {American Physical Society},
  doi = {10.1103/PhysRevLett.121.123601},
  url = {https://link.aps.org/doi/10.1103/PhysRevLett.121.123601}
}

@article{Zanner2022,
  title = {Coherent control of a multi-qubit dark state in waveguide quantum electrodynamics},
  author = {Zanner,Maximilian  and Orell, Tuure and Schneider, Christian M. F. and Albert, Romain and Oleschko, Stefan and Juan, Mathieu L. and Silveri, Matti and Kirchmair, Gerhard
},
  journal = {Nature Physics},
  volume = {18},
  issue = {5},
  pages = {538-543},
  numpages = {6},
  year = {2022},
  month = {Mar},
  doi = {10.1038/s41567-022-01527-w},
  url = {https://doi.org/10.1038/s41567-022-01527-w}
}

@article{Zanner2025,
  title = {Spatial addressing of qubits in a dispersive waveguide},
  author = {Zanner, Maximilian and Albert, Romain and Rosenthal, Eric I. and Casulleras, Silvia and Yang, Ian and Schneider, Christian M.F. and Romero-Isart, Oriol and Kirchmair, Gerhard},
  journal = {Phys. Rev. Appl.},
  volume = {24},
  issue = {1},
  pages = {014051},
  numpages = {16},
  year = {2025},
  month = {Jul},
  publisher = {American Physical Society},
  doi = {10.1103/np2t-48rm},
  url = {https://link.aps.org/doi/10.1103/np2t-48rm}
}

@article{Kannan2023,
  title = {On-demand directional microwave photon emission using waveguide quantum electrodynamics
},
  author = {Kannan, Bharath and Almanakly, Aziza and Sung, Youngkyu and Di Paolo, Agustin
and Rower, David A.
and Braumüller, Jochen
and Melville, Alexander
and Niedzielski, Bethany M.
and Karamlou, Amir
and Serniak, Kyle
and Vepsäläinen, Antti
and Schwartz, Mollie E.
and Yoder, Jonilyn L.
and Winik, Roni and Wang, Joel I-Jan and Orlando, Terry P. and Gustavsson, Simon and Grover, Jeffrey A.
and Oliver, William D.
},
  journal = {Nature Physics},
  volume = {19},
  issue = {3},
  pages = {394-400},
  numpages = {7},
  year = {2023},
  month = {Jan},
  doi = {10.1038/s41567-022-01869-5},
  url = {https://doi.org/10.1038/s41567-022-01869-5
}}

@InProceedings{Kockum2021,
author="Frisk Kockum, Anton",
editor="Takagi, Tsuyoshi
and Wakayama, Masato
and Tanaka, Keisuke
and Kunihiro, Noboru
and Kimoto, Kazufumi
and Ikematsu, Yasuhiko",
title="Quantum Optics with Giant Atoms---the First Five Years",
booktitle="International Symposium on Mathematics, Quantum Theory, and Cryptography",
year="2021",
publisher="Springer Singapore",
address="Singapore",
pages="125--146",
isbn="978-981-15-5191-8",
doi={10.1007/978-981-15-5191-8_12}
}

\end{document}